\documentclass[10pt,a4paper]{article}

\usepackage[T1]{fontenc}
\usepackage[utf8]{inputenc}
\usepackage{amssymb,amsmath,amsthm,amsfonts}    
\usepackage{bm}                                 
\usepackage{mathtools}                          
\usepackage{stmaryrd}                           
\usepackage{breqn}                              
\usepackage[textfont=footnotesize,labelfont={footnotesize,bf}]{caption}
\usepackage{subcaption}
\usepackage{booktabs}
\usepackage[inline]{enumitem}
\usepackage{graphicx}
\usepackage[dvipsnames]{xcolor}
\usepackage[top=30mm,right=30mm,left=30mm,bottom=30mm]{geometry}
\usepackage[colorlinks=true,linkcolor=blue,urlcolor=blue]{hyperref}
\usepackage[affil-it,auth-sc]{authblk}
\usepackage[colorinlistoftodos,textwidth=25mm,textsize=footnotesize]{todonotes}
\usepackage{lipsum}                             
\usepackage{circledsteps}
\usepackage[
	backend=biber,
	style=numeric,
	citestyle=numeric-comp,
	maxbibnames=99,
	minbibnames=99,
	maxcitenames=2,
	mincitenames=1,
	giveninits=true,
	sorting=none,
	sortlocale=auto,
	natbib=true,
	url=false,
	isbn=false,
	doi=true,
	eprint=true
]{biblatex}
\newcommand{\bzero}{\bm 0}
\newcommand{\ba}{\bm a}
\newcommand{\bb}{\bm b}

\newcommand{\be}{\bm e}
\newcommand{\bef}{\bm f}
\newcommand{\bg}{\bm g}

\newcommand{\bk}{\bm k}

\newcommand{\bn}{\bm n}

\newcommand{\bp}{\bm p}
\newcommand{\bq}{\bm q}

\newcommand{\bv}{\bm v}

\newcommand{\bx}{\bm x}

\newcommand{\bA}{\bm A}

\newcommand{\bD}{\bm D}

\newcommand{\bF}{\bm F}

\newcommand{\bI}{\bm I}

\newcommand{\bK}{\bm K}
\newcommand{\bL}{\bm L}

\newcommand{\bP}{\bm P}

\newcommand{\bS}{\bm S}
\newcommand{\bT}{\bm T}

\newcommand{\bW}{\bm W}

\newcommand{\bZ}{\bm Z}

\newcommand{\fC}{\mathbb C}

\newcommand{\fR}{\mathbb R}

\newcommand{\fZ}{\mathbb Z}

\newcommand{\mB}{\mathcal B}
\newcommand{\mC}{\mathcal C}

\newcommand{\mE}{\mathcal E}

\newcommand{\mL}{\mathcal L}

\graphicspath{{./}{./figures/}}                 

\title{Material instability and subsequent restabilization from homogenization of periodic elastic lattices}

\author[1]{Davide Bigoni\footnote{Corresponding author: e-mail: \href{mailto:bigoni@ing.unitn.it}{bigoni@ing.unitn.it}; phone: +39\,0461\,282507.}}
\author[1]{Andrea Piccolroaz}

\affil[1]{Department of Civil, Environmental, and Mechanical Engineering, University of Trento, Trento, Italy}

\date{\today}

\begin{document}

\def\gdp{\makebox{\raisebox{-.215ex}{$\Box$}\hspace{-.778em}$\times$}}

\maketitle

\begin{abstract}
\noindent
Two classes of non-linear elastic materials are derived via two-dimensional homogenization. These materials are equivalent to a periodic grid of axially-deformable and axially-preloaded structural elements, subject to incremental deformations that involve bending, shear, and normal forces. The unit cell of one class is characterized by elements where deformations are lumped within a finite-degrees-of-freedom framework. In contrast, the other class involves smeared deformation, modelled as flexurally deformable rods with sufficiently high axial compliance. Under increasing compressive load, the elasticity tensor of the equivalent material loses positive definiteness and subsequently undergoes an ellipticity loss. Remarkably, in certain conditions, this loss of stability is followed by a subsequent restabilization; that is, the material re-enters the elliptic regime and even the positive definiteness domain and simultaneously, the underlying elastic lattice returns to a stable state. This effect is closely related to the axial compliance of the elements. 

The lumped structural model is homogenized using a purely mechanical approach (whose results are also confirmed via formal homogenization based on variational calculus), resulting in an analytical closed-form solution that serves as a reference model. 
Despite its simplicity, the model exhibits a surprisingly rich mechanical behaviour. Specifically, for certain radial paths in stress space: (i.) stability is always preserved; (ii.) compaction, shear, and mixed-mode localization bands emerge; (iii.) shear bands initially form, but later ellipticity is recovered, and finally, mixed-mode localization terminates the path. This lumped structural model is (at least in principle) realizable in practice and offers an unprecedented and vivid representation of strain localization modes, where the corresponding equations remain fully \lq manageable by hand'. The structural model with smeared deformability behaves similarly to the discrete model but introduces a key distinction: \lq islands' of instability emerge within a broad zone of stability. This unique feature leads to unexpected behaviour, where shear bands appear, vanish and reappear along radial stress paths originating from the unloaded state. 

Our results: (i.) demonstrate new possibilities for exploiting structural elements within the elastic range,  characterized by a finite number of degrees of freedom, to create architected materials with tuneable instabilities, (ii.) introduce reconfigurable materials characterized by \lq islands' of stability or instability.
\end{abstract}

\paragraph{Keywords}
Lattice materials \textperiodcentered\
Material instability \textperiodcentered\
Homogenization

\section{Introduction}
\label{sec:introduction}
Sufficiently high axial compliance has been shown to induce restabilization in an elastic rod susceptible to Euler buckling \cite{tarnai_destabilizing_1980, feodosyev_selected_1977, potier-ferry_foundations_1987, koutsogiannakis_double_2023}. The basic mechanics behind this behaviour is simple: the critical load for the bifurcation of an axially-compressed rod is inversely proportional to the square of its length. Thus, axial deformation reduces the rod's length, resulting in a stabilizing effect. Rods undergoing significant length variation have attracted renewed interest, especially in deployable structures (for instance, booms, antennas, and solar arrays spacecraft \cite{pellegrino_deployable_2001}) and soft robot arms, where configurational forces often play a role \cite{bigoni_instability_2014}. 

\begin{figure}[hbt!]
    \centering
    \includegraphics[width=0.75\linewidth]{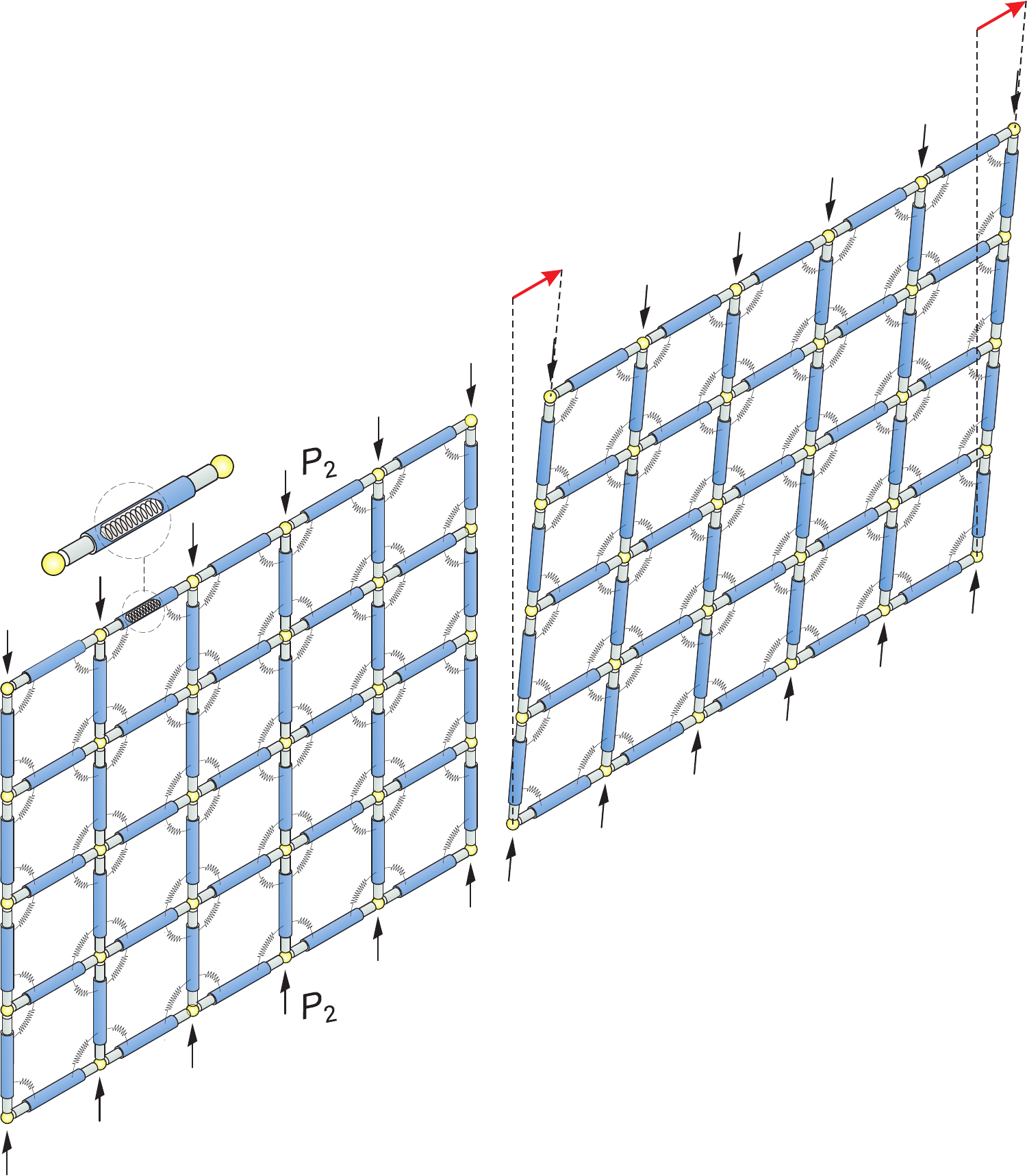}
    \caption{The proof-of-concept model (left) is an architected material showing restabilization, when subject to increasing compressive prestress. Restabilization occurs both in the structure and in an equivalent elastic continuum. The structural elements are axially deformable, but flexurally rigid, and are connected through elastic hinges, permitting shear deformation (right).}
    \label{maglia_rigida}
\end{figure}

This article extends the concept of restabilization, previously known only in structural mechanics, to elastic {\it solids}. The proof-of-concept model, illustrated in Fig.~\ref{maglia_rigida}, demonstrates this idea. The model consists of a rectangular lattice made up of axially deformable but otherwise rigid bars, connected by elastic hinges that condense the shear deformability. 
%

The lattice is doubly periodic, infinite in extent, and preloaded with vertical and horizontal forces, leading to a state where only axial internal forces are present. A global bifurcation corresponding to a shear band mode occurs at sufficiently high compressive loads. This lattice serves as a conceptual model for an architected material that can be made equivalent to an elastic, prestressed solid. 
The explicit determination of the equivalent elasticity tensor (performed via purely mechanical considerations and confirmed through rigorous homogenization based on variational calculus), characterizing the incremental response of the material equivalent to the lattice for every combination of biaxial prestress, enables the analysis of stability criteria such as positive definiteness (PD), strong ellipticity (SE), and ellipticity (E). 
These analyses indisputably demonstrate instability followed by restabilization within certain regions of the stability map, revealing \lq islands of stability in a sea of instability'.

Surprisingly, the essential mechanical model illustrated in Fig.~\ref{maglia_rigida} reveals a rich mechanical behaviour and provides a vivid example of ellipticity loss, which is {\it per se} interesting, showing the emergence of shear band modes, compaction bands, and mixed-mode localizations. 

\begin{figure}[hbt!]
    \centering
    \includegraphics[width=0.98\linewidth]{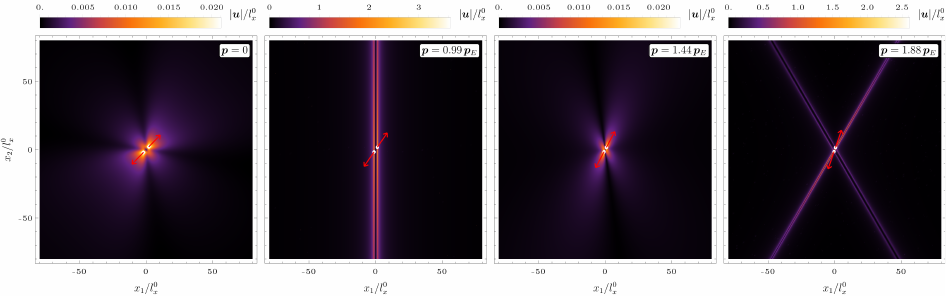}
    \caption{Displacement maps at radially increasing compressive prestress $p$, along a radial path emanating from the unloaded state. The maps are obtained by applying a perturbative dipole to an elastic material equivalent to a square grid of axially prestressed elastic rods (see Section \ref{ritornano} for the details). 
    From left to right: (i.) initially, shear bands are not present, and the (SE) condition holds; (ii.) a vertical shear band occurs when the (SE)-(Parabolic) boundary is grazed; (iii.) the shear band disappears, and the material is fully restabilized; (iv.) finally, two inclined shear bands emerge when the (SE)-(Hyperbolic) boundary is approached. Note that the shear bands are approached two times while the entire stress path remains inside the (SE) regime.}
    \label{figata}
\end{figure}

To ensure these findings are not limited to discrete lattices with a finite number of degrees of freedom, the study is complemented by a demonstration of restabilization through the homogenization of a periodic lattice comprising axially and flexurally deformable elastic rods. This approach extends previous results \cite{bordiga_dynamics_2021} to include substantial axial compliance. The analysis of this new structure reveals two novel and unexpected features: (i) islands of instability now emerge within a stable zone, and consequently, (ii) shear bands may {\it appear, disappear, and finally reappear along a radial stress path originating from the null stress state}. Fig.~\ref{figata} illustrate this phenomenon: starting from a stable configuration and following a compressive radial path, deformation localization initially emerges as a vertical shear band near $p_E$, corresponding to the first loss of (SE); as the compressive load increases beyond $p_E$, the material re-enters a stable regime, causing the shear band to fade; further compression induces strain localization to reappear in the form of a mixed mode.
%

It is important to note that the restabilization observed in the present article differs from that reported in \cite{bordiga_tensile_2022}. The latter reflects a sort of \lq discrepancy' in the homogenization result, where the solid recovers stability while the underlying lattice remains unstable. 
This phenomenon is related to the well-established fact that (SE) in the equivalent solid is unaffected by the occurrence of microscopic bifurcations in the lattice \cite{geymonat_homogenization_1993,triantafyllidis_comparison_1993,lopez-pamies_overall_2006}.  

Our results reveal a complex interplay of instabilities and offer insights for designing materials where stability loss and recovery can be explicitly engineered while maintaining a clear mechanical interpretation. This suggests the possibility of developing materials capable of operating in a stable regime at loads exceeding the first critical load in a vein similar to supercritical driveshafts functioning at speeds higher than their natural critical frequency.


The article is organized as follows. 
Section \ref{modellazzo} introduces a straightforward technique for homogenizing the proof-of-principle structure shown in Fig.~\ref{maglia_rigida}, determining the equivalent elastic solid. For this solid, the (PD) and (SE) stability conditions are also derived.
Section \ref{discrcr} presents a comprehensive analysis of the unit cell, viewed as a repeating unit that spans the entire plane and generates the periodic structure. The unit cell is subjected to general displacement fields. The equivalent of the (PD) condition for the solid is obtained for the elastic grid, and the two conditions are shown to coincide.
A stricter exclusion condition than (PD) is also derived for the discrete structure under uniform prestress, using the Floquet-Bloch wave representation of the displacement field. This condition reduces to the (SE) condition for the solid when considering infinite wavelengths, thereby highlighting both the strengths and limitations of the homogenization process.
A formal homogenization algorithm, based on variational calculus and Gamma-convergence, is presented in Section \ref{gammastokk}, demonstrating that the results derived from purely mechanical principles are rigorous. More generally, the formal proof leads to the acoustic tensor of the homogenized material and is thus analogous to the dynamic asymptotic homogenization technique developed in \cite{bordiga_dynamics_2021}.
Section \ref{esempioni} provides examples demonstrating restabilization in terms of both (PD) and (SE). These examples illustrate scenarios where the strongly elliptic boundary is punctured up to three times, documenting the occurrence of shear, compaction bands and mixed-mode localization.
Finally, Section \ref{isolotti} concludes the article by introducing a lattice composed of axially and flexurally deformable elastic rods. The homogenization of this lattice and its related results are discussed, showcasing the phenomena of 'instability islands,' along with the emergence, disappearance, and reemergence of localization.

\section{Homogenization of the proof-of-concept model} 
\label{modellazzo}
\begin{figure}[hbt!]
    \centering
    \includegraphics[width=0.8\linewidth]{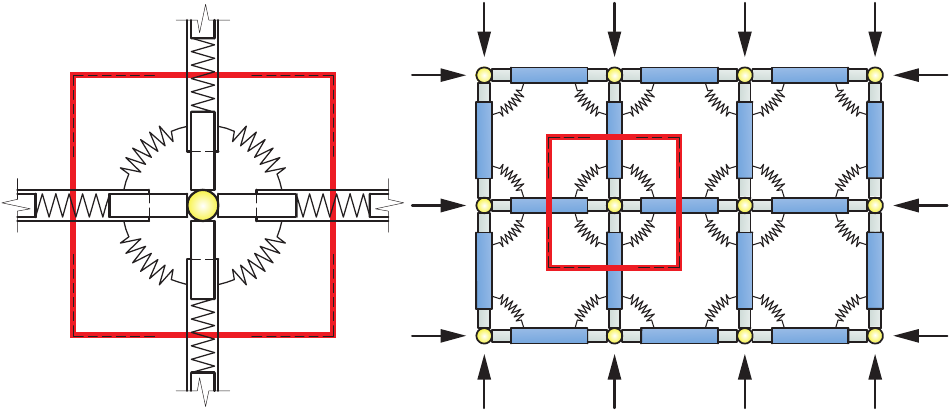}
    \caption{Front view of Fig.~\ref{maglia_rigida}, showing the unit cell and a portion of the lattice that forms the proof-of-concept model showing restabilization of an equivalent elastic continuum when subject to increasing compressive prestress. The elements are axially deformable but otherwise rigid and are interconnected through elastic hinges.}
    \label{maglia_rigida_2}
\end{figure}

A proof-of-concept mechanical model is established, consisting of axially deformable (with stiffnesses $k_x$ and $k_y$) but flexurally rigid elements that form the rectangular (periodic and infinite) grid shown in Fig.~\ref{maglia_rigida} in an axonometric view and Fig.~\ref{maglia_rigida_2} in a front view, with the unit cell highlighted. The bars, with lengths $l^0_x$ and $l^0_y$ in the undeformed configuration, are elastically hinged together (with stiffness $k_o$). 
%
%

\subsection{Homogeneous deformation of the grid and the equivalent continuum} 

\begin{figure}[hbt!]
    \centering
    \includegraphics[width=0.5\linewidth]{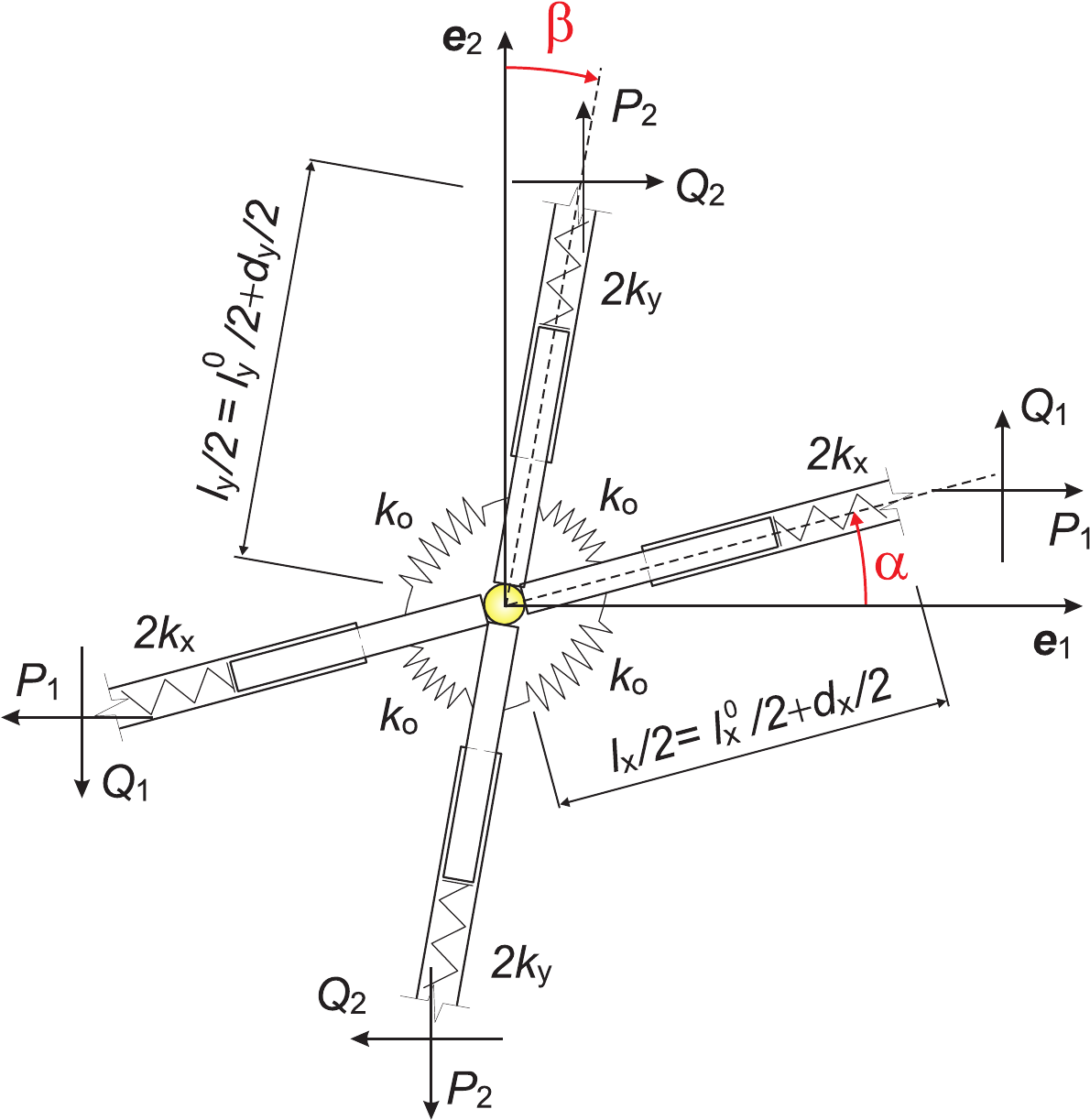}
    \caption{The deformation of the unit cell, associated with a linear displacement field, which corresponds to a uniform strain used for homogenization. Note that the stiffnesses pertaining to one half of a longitudinal spring are $2k_x$ and $2k_y$.}
    \label{maglia_rigida_3}
\end{figure}

For any linear displacement field, the grid is constrained to deform uniformly, reducing its degrees of freedom from 10 to 4: the two elongations of horizontal and vertical bars, $d_1 = d_3 = d_x$, and $d_2 = d_4 = d_y$, along with two inclination angles, $\alpha_1 = \alpha_3 = \alpha$, and $\alpha_2 = \alpha_4 = -\beta$. 
In the reference system $\be_1$--$\be_2$, the deformed configuration of the unit cell, when subject to vertical and horizontal forces $P_1$, $Q_1$ and $P_2$, $Q_2$, is shown in Fig.~\ref{maglia_rigida_3}. The half-lengths of the bars defining the unit cell become $l_x/2$ and $l_y/2$, satisfying
\begin{equation}
    l_x = l_x^0 + d_x, 
    \quad
    l_y = l_y^0 + d_y.  
\end{equation}
%

Noting that the half-springs pertaining to the unit cell have stiffnesses $2 k_x$ and $2 k_y$ in the horizontal and vertical bars, respectively, the total potential energy in the deformed configuration is 
\begin{multline}
    W = 
    2k_o (\alpha+\beta)^2 + 2k_x\left(\frac{d_x}{2}\right)^2 + 2k_y\left(\frac{d_y}{2}\right)^2 \\
    + P_1l^0_x \left[1-\left(1+\frac{d_x}{l^0_x} \right) \cos\alpha \right] 
    + P_2l^0_y \left[1-\left(1+\frac{d_y}{l^0_y} \right) \cos\beta \right] \\
    - Q_1l^0_x \left(1+\frac{d_x}{l^0_x} \right) \sin\alpha
    - Q_2l^0_y \left(1+\frac{d_y}{l^0_y} \right) \sin\beta .
\end{multline}

\subsubsection{Equilibrium}
The stationarity of $W$ with respect to $\alpha$, $\beta$, $d_x$, and $d_y$ leads to the equilibrium equations 
\begin{equation}
\label{peretta}
    \begin{aligned}
        & 4\frac{k_o}{l^0_x}(\alpha+\beta) + P_1\left(1+\frac{d_x}{l^0_x}\right)\sin\alpha - Q_1\left(1+\frac{d_x}{l^0_x}\right)\cos\alpha = 0 , \\[3mm]
        & 4\frac{k_o}{l^0_y}(\alpha+\beta) + P_2\left(1+\frac{d_y}{l^0_y}\right)\sin\beta - Q_2\left(1+\frac{d_y}{l^0_y}\right)\cos\beta = 0 , \\[3mm]
        & k_x d_x - P_1\cos\alpha - Q_1\sin\alpha = 0 , \\[3mm]
        & k_y d_y - P_2\cos\beta - Q_2\sin\beta = 0 .
    \end{aligned}
\end{equation}
With reference to a unit cell enclosing the deformed configuration shown in Fig.~\ref{maglia_rigida_3}, the Cauchy, $T_{ij}$, and Piola-Kirchhoff, $S_{ij}$, stresses can be introduced as 
\begin{equation}
    \begin{aligned}
        T_{11} &= \frac{P_1}{l_y},
        &T_{12} &= \frac{Q_2}{l_x},
        &T_{21} &= \frac{Q_1}{l_y},
        &T_{22} &= \frac{P_2}{l_x}, \\
        S_{11} &= \frac{P_1}{l_y^0},
        &S_{12}  &= \frac{Q_2}{l_x^0},
        &S_{21} &= \frac{Q_1}{l_y^0},
        &S_{22} &= \frac{P_2}{l_x^0},
    \end{aligned}
\end{equation}
and the equilibrium equations \eqref{peretta} can be rewritten as 
\begin{equation}
\label{peretta99}
    \begin{aligned}
        &4\frac{k_o}{l^0_xl^0_y}(\alpha+\beta) + S_{11} \left(1+\frac{d_x}{l^0_x} \right) \sin\alpha - S_{21} \left(1+\frac{d_x}{l^0_x} \right) \cos\alpha = 0, \\[3mm]
        &4\frac{k_o}{l^0_xl^0_y}(\alpha+\beta) + S_{22}  \left(1+\frac{d_y}{l^0_y} \right)\sin\beta - S_{12} \left(1+\frac{d_y}{l^0_y} \right) \cos\beta = 0, \\[3mm]
        &\frac{k_x d_x}{l^0_y} - S_{11}\cos\alpha - S_{21}\sin\alpha = 0, \\[3mm]
        &\frac{k_y d_y}{l^0_x} - S_{22}\cos\beta - S_{12}\sin\beta = 0.
    \end{aligned}
\end{equation}
It is assumed that $\alpha$, $\beta$, $d_x$, $d_y$, $T_{ij}$, $S_{ij}$, and thus also $\lambda_x$ and $\lambda_y$ are all functions of a time-like parameter governing the deformation. 
Moreover, a biaxially-deformed configuration in which the bars remain aligned parallel to their initial position is postulated, so that $\alpha=\beta=0$, $Q_1=Q_2=0$, thus $S_{12}=S_{21}=0$. 
In this situation, equations (\ref{peretta99}) allow to relate the stretches 
\begin{equation}
\label{straz}
    \lambda_1 = \frac{l_x}{l_x^0} , 
    \quad
    \lambda_2 = \frac{l_y}{l_y^0} , 
\end{equation}
to the prestress as
\begin{equation}
\label{streccioni}
    \lambda_1 = 1 + \frac{S_{11}l^0_y}{k_xl^0_x} ,
    \quad
    \lambda_2 = 1 + \frac{S_{22}l^0_x}{k_yl^0_y} .
\end{equation}
The continuum equivalent to the biaxially stretched discrete structure is subject to a uniform prestretch characterized by the principal stretches $\lambda_1$ and $\lambda_2$, eqns.~\eqref{straz}, so that the deformation gradient $\bF$ and its determinant $J$ are
\begin{equation}
    [\bF] = 
    \begin{bmatrix}
        \lambda_1 &  0 \\
        0 & \lambda_2
    \end{bmatrix},
    \quad 
    J = \lambda_1\lambda_2,
\end{equation}
and the validity of the relation $\bS = J \bT\bF^{-T}$ can easily be checked.

\subsubsection{Incremental equilibrium}
The incremental equilibrium equations corresponding to eqs.~\eqref{peretta99} are
\begin{equation}
\label{lamarianna22}
    \begin{aligned}
        S'_{11} &= \frac{k_x\dot{d}_x}{l_y^0},
        &S'_{12} &= \frac{4k_o}{l_x^0l_y^0} \frac{\dot{\alpha}+\dot{\beta}}{\lambda_2} + S_{22} \dot{\beta}, \\[3mm]
        S'_{21} &= \frac{4k_o}{l_x^0l_y^0} \frac{\dot{\alpha}+\dot{\beta}}{\lambda_1} + S_{11} \dot{\alpha}, 
        &S'_{22} &= \frac{k_y\dot{d}_y}{l_x^0},
    \end{aligned}
\end{equation}
where a superimposed dot denotes increments, while an increment in $\bS$ is indicated with a dash, $S'_{ij}$. It should be noted from eq.~\eqref{lamarianna22} that $S'_{12}$ is different from $S'_{21}$, except when the prestress is null, $S_{11}=S_{22}=0$.

The incremental equations relating incremental nominal stresses $S'_{ij}$ to the incremental deformation governed by $\dot{\alpha}$, $\dot{\beta}$, $\dot{d}_x$, $\dot{d}_y$ can also be expressed as functions of the sole  prestress components $S_{11}$ and $S_{22}$ in the form
\begin{equation}
\label{lamarianna}
    \begin{aligned}
        S'_{11} &= \frac{k_x\dot{d}_x}{l_y^0},
        &S'_{12} &= \frac{4k_o}{l_x^0l_y^0} \frac{\dot{\alpha}+\dot{\beta}}{1+\frac{S_{22}l^0_x}{k_yl^0_y}} + S_{22} \dot{\beta}, \\[3mm]
        S'_{21} &= \frac{4k_o}{l_x^0l_y^0} \frac{\dot{\alpha}+\dot{\beta}}{1+\frac{S_{11}l^0_y}{k_xl^0_x}} + S_{11} \dot{\alpha},
        &S'_{22} &= \frac{k_y\dot{d}_y}{l_x^0}. 
    \end{aligned}
\end{equation}
A prestressed elastic solid, governed by an incremental constitutive equation relating the increment of the first Piola-Kirchhoff stress $S'_{ij}$ to the incremental deformation gradient $F'_{h,k}$, 
\begin{equation}
\label{incre33}
    S'_{ij} = \mathbb{G}_{ijhk} F'_{hk}, 
\end{equation}
can be made equivalent to the incremental behaviour of the grid, equations \eqref{lamarianna}, by observing the kinematic equivalence between continuum and lattice 
\begin{equation}
\begin{array}{llllll}
    \displaystyle
    \quad
    F_{11}= \frac{l_x}{l^0_x}, &
    \displaystyle
    \quad
    F_{12} = 0, &
    \displaystyle
    \quad
    v_{1,1} = \frac{\dot{d}_x}{l_x}, &
    \displaystyle
    \quad
    v_{1,2} = \dot{\beta}, & 
    \displaystyle
    \quad
    F'_{11} = 
    \frac{\dot{d}_x}{l^0_x} & 
    \displaystyle
    \quad
    F'_{12} = \dot{\beta}\frac{l_y}{l_y^0}, \\[5mm]
    \displaystyle
    \quad
    F_{21}=0, &
    \displaystyle
    \quad
    F_{22} = \frac{l_y}{l^0_y}, &
    \displaystyle
    \quad
    v_{2,1} = \dot{\alpha}, & 
    \displaystyle
    \quad
    v_{2,2} = \frac{\dot{d}_y}{l_y}, &
    \displaystyle
    \quad
    F'_{21} = \dot{\alpha} \frac{l_x}{l_x^0}, &
    \displaystyle
    \quad
    F'_{22} =  \frac{\dot{d}_y}{l^0_y}, 
    \end{array}
\end{equation}
where the gradient of incremental deformation $v_{i,j}$ satisfies $F'_{ik}=v_{i,j}F_{jk}$. 
Therefore, the elastic fourth-order tensor of the solid equivalent to the grid (only the non-null components are reported) becomes 
\begin{equation}
\begin{array}{lll}
\displaystyle
    \mathbb{G}_{1111} = \frac{k_x l^0_x}{l^0_y}, & 
    \displaystyle 
    \mathbb{G}_{2222} = \frac{k_y l^0_y}{l^0_x}, \\[5mm]
    \displaystyle
    \mathbb{G}_{1212} = \frac{4k_o}{l_x^0l_y^0\lambda_2^2} + \frac{S_{22}}{\lambda_2}, & 
    \displaystyle 
    \mathbb{G}_{1221} = \frac{4k_o}{l_x^0l_y^0\lambda_1\lambda_2}, \\[5mm]
    \displaystyle
    \mathbb{G}_{2112} = \frac{4k_o}{l_x^0l_y^0\lambda_1\lambda_2}, & 
    \displaystyle 
    \mathbb{G}_{2121} = \frac{4k_o}{l_x^0l^0_y\lambda_1^2} +\frac{S_{11}}{\lambda_1}, 
\end{array}
\end{equation}
where the major symmetry $\mathbb{G}_{2112}=\mathbb{G}_{1221}$ should be noted.

On the introduction of the Oldroyd derivative of the Kirchhoff stress, $\bK=J\bT$, as
\begin{equation}
    \stackrel{\circ}{\bK} = \bS'\bF^T - (\nabla \bv) \bK, 
\end{equation}
it is found that $\stackrel{\circ}{\bK}$ is related to the Eulerian strain increment $\bD = (\nabla \bv + \nabla \bv^T)/2$ by 
\begin{equation} 
    \stackrel{\circ}{\bK} = \mathbb{H} [\bD], 
\end{equation}
where
\begin{equation}
\label{oldoldold}
    \mathbb{H}_{1111} = \frac{k_x l_x}{l^0_y}, 
    \quad
    \mathbb{H}_{2222} = \frac{k_y l_y}{l^0_x},
    \quad
    \mathbb{H}_{1212} = \mathbb{H}_{2112} = 
    \mathbb{H}_{2121} = \mathbb{H}_{1221} = \frac{8k_o}{l^0_xl^0_y}, 
\end{equation}
so that tensor $\mathbb{H}$ possesses both the major and the minor symmetries. If the bars remain of finite length during the deformation (as it is imposed by elementary considerations), all components (\ref{oldoldold}) are strictly positive and never vanish. The constitutive tensor $\mathbb{H}$ defines an orthotropic material and is positive definite.

\subsubsection{Relative Lagrangean description}

In a {\it relative Lagrangian description}, in which the current configuration is assumed as a reference, the constitutive response can be written in terms of the incremental first Piola-Kirchhoff stress $\dot{S}_{ij}$ as
\begin{equation}
    \dot{S}_{ij} = \mathcal{E}_{ijhk} v_{h,k}, 
\end{equation}
where 
\begin{equation}
    \dot{S}_{11} = S'_{11} \frac{l_y^0}{l_y}, \quad
    \dot{S}_{22} = S'_{22} \frac{l_x^0}{l_x}, \quad
    \dot{S}_{12} = S'_{12} \frac{l_x^0}{l_x}, \quad
    \dot{S}_{21} = S'_{21} \frac{l_y^0}{l_y}, 
\end{equation}
so that the relationships between the two introduced elasticity tensors are
\begin{equation}
\label{polpettona}
\begin{aligned}
    \mathcal{E}_{1111} &= \mathbb{G}_{1111}\frac{l_xl_y^0}{l^0_xl_y} , & 
    \mathcal{E}_{2222} &= \mathbb{G}_{2222}\frac{l_x^0l_y}{l_xl_y^0} , \\
    \mathcal{E}_{1212} &= \mathbb{G}_{1212}\frac{l_x^0l_y}{l_xl^0_y} , & 
    \mathcal{E}_{1221} &= \mathbb{G}_{1221} \\
    \mathcal{E}_{2112} &= \mathbb{G}_{2112} , & 
    \mathcal{E}_{2121} &= \mathbb{G}_{2121}\frac{l_xl^0_y}{l^0_xl_y} , 
\end{aligned}
\end{equation}
and therefore the only non-null components of tensor $\mathcal{E}$ can be written as
\begin{equation}
\label{paola_gatti}
\begin{aligned}
    \displaystyle
    \mathcal{E}_{1111} &= \frac{k_x l_x}{l_y}, & 
    \mathcal{E}_{2222} &= \frac{k_y l_y}{l_x}, \\
    \mathcal{E}_{1212} &= \frac{4k_o}{l_xl_y} + T_{22}, & 
    \mathcal{E}_{1221} &= \frac{4k_o}{l_xl_y}, \\ 
    \mathcal{E}_{2112} &= \frac{4k_o}{l_xl_y}, & 
    \mathcal{E}_{2121} &= \frac{4k_o}{l_xl_y} + T_{11}.
\end{aligned}
\end{equation}

In a relative Lagrangian description, the Oldroyd derivative of the Kirchhoff stress equals the material time derivative of the second Piola-Kirchhoff stress, $\bT^{(2)} = \bF^{-1}\bS$. Therefore, in a relative Lagrangian description, it is possible to define a tensor $\mathcal{H}$ such that
\begin{equation}
    \dot{\bT}^{(2)} = \mathcal{H} [\bD],
\end{equation}
with 
\begin{equation}
\label{paolona_gattona}
    \mathcal{H}_{1111} = \frac{k_x l_x^0}{l_y} ,  
    \quad
    \mathcal{H}_{2222} = \frac{k_y l_y^0}{l_x} , 
    \quad
    \mathcal{H}_{1212} = \mathcal{H}_{1221} = \mathcal{H}_{2112} = \mathcal{H}_{2121} = \frac{4k_o}{l_xl_y} .
\end{equation}
Consequently, the tensor $\mathcal{E}$ can be written as 
\begin{equation}
    \mathcal{E} = \mathcal{H} + \bI \gdp \bT, 
\end{equation}
where the tensorial product \lq $\gdp$' is defined in such a way that for every tensor $\bL$, it is $(\bI\gdp\bT)\bL=\bL\bT^T=\bL\bT$.

\subsubsection{The bifurcation exclusion condition: positive definiteness (PD) of the constitutive tensor $\mathcal{E}$}

Equations (\ref{paolona_gattona}) show that:

\begin{itemize}

\item The tensor $\mathcal{H}$ is positive definite if and only if all springs have strictly positive stiffness.  

\item Positive definiteness of the tensor $\mathcal{E}$, called (PD), namely, $\bL \cdot \mathcal{E} \bL >0$, holding for every non-null $\bL = \bD+\bW$ ($\bW$ being the spin tensor), can be written as 
\begin{equation}
    \mathcal{E}_{1111} D_{11}^2 + \mathcal{E}_{2222} D_{22}^2 + 4\mathcal{H}_{1212} D_{12}^2 
    + (T_1+T_2) D_{12}^2 + (T_1+T_2) W_{12}^2 - 2(T_{1}-T_{2})D_{12}W_{12} > 0,
\end{equation}    
valid for every component $D_{ij}$ and $W_{12}$, not identically null. (PD) is a condition that, when holding at every point of a solid, excludes bifurcation of the incremental solution for any mixed boundary condition \cite{bigoni_nonlinear_2012}.

Due to the fact that $\bD$ and $\bW$ are independent tensors and each of them can be set to be null independently of the other, $T_1+T_2>0$ is a necessary condition for positive definiteness of $\mathcal{E}$, together with other two trivial conditions, $\mathcal{E}_{1111}>0$ and $\mathcal{E}_{2222}>0$. The interaction between incremental shear strain and spin is more intricate and, at fixed $D_{12}$ and for $T_1+T_2>0$, requires consideration of the function 
\begin{equation}
    z(W_{12}) = (T_1+T_2) W_{12}^2 - 2(T_{1}-T_{2})D_{12}W_{12}, 
\end{equation}
which reaches its minimum at 
\begin{equation}
    \tilde{W}_{12} = \frac{T_1-T_2} {T_1+T_2}D_{12}. 
\end{equation}

Therefore, the inequality 
\begin{equation}
    \bL \cdot \mathcal{E} \bL \geq 
    \mathcal{E}_{1111} D_{11}^2 + \mathcal{E}_{2222} D_{22}^2 + 4\left(\mathcal{H}_{1212} + \frac{T_1T_2}{T_1+T_2}\right) D_{12}^2,
\end{equation}
leads to the following necessary and sufficient conditions for the positive definiteness of $\mathcal{E}$, the so-called (PD) conditions, 
\begin{equation}
\label{pd}
    \mathcal{H}_{1111}+T_1>0, \quad
    \mathcal{H}_{2222}+T_2>0, \quad
     T_1+T_2 > 0, \quad
    \mathcal{H}_{1212} + \frac{T_1T_2}{T_1+T_2} > 0 \quad
    \Longleftrightarrow \quad \text{(PD)} 
\end{equation}
ruling out the possibility of any bifurcation when holding at every point of a stressed body. Note that: (i.)  conditions (\ref{pd})$_1$ and (\ref{pd})$_2$ must be trivially satisfied, as violating either of them would imply either the annihilation of a bar or the length becoming negative; (ii.) condition (\ref{pd})$_3$ is purely geometric and does not involve the material response; (iii.) all conditions (\ref{pd}) are trivially satisfied for both positive $T_1$ and $T_2$, so that any instability is ruled out when the stresses are both tensile. 

The four eigenvalues of the tensor $\mE$ are: 
\begin{equation}
    \frac{k_x l_x}{l_y}, \quad \frac{k_y l_y}{l_x}, \quad
    \frac{4 k_o}{l_x l_y} + \frac{T_1 + T_2}{2} \pm \sqrt{\frac{16 k_o^2}{l_x^2 l_y^2} + \left(\frac{T_1 - T_2}{2}\right)^2}, 
\end{equation}
and thus, the (PD) condition, Eq.~(\ref{pd}), can be rewritten in the equivalent form  
\begin{equation}
\label{orchite}
     l_x>0, \quad  l_y>0, \quad
    \frac{4 k_o}{l_x l_y} + \frac{T_1 + T_2}{2} - \sqrt{\frac{16 k_o^2}{l_x^2 l_y^2} + \left(\frac{T_1 - T_2}{2}\right)^2} > 0 
    \quad
    \Longleftrightarrow \quad \text{(PD)} 
\end{equation}

\item The tensor $\mathbb{H}$ is positive definite on the restriction to symmetric tensors if and only if all springs have strictly positive stiffness. 

\item Recalling the equation $F'_{ik}=v_{i,j}F_{jk}$, the positive definiteness of the tensor $\mathbb{G}$, equations (\ref{polpettona}), can easily be shown to be equivalent to (PD), equation (\ref{pd}).

\end{itemize}

At a fixed value of $P_2>0$, $l_y=l_y^0+P_2/l_y$ is also fixed, so that $l_yP_2>0$; moreover, $T_1+T_2$ is assumed to be strictly positive, which implies $P_1l_x+P_2l_y>0$. Under these assumptions, the vanishing of an eigenvalue of $\mE$ occurs at
\begin{equation}
\label{perona}
    \displaystyle 
    \frac{P_1}{k_xl^0_x} = 
    -\frac{1}{2} \left(1 \pm \sqrt{1-\frac{16 k_o l_yP_2}{k_x(l_x^0)^2(4k_o+l_yP_2)}}\,\right), 
\end{equation}
where the \lq $\pm$' shows the {\it possibility}  of restabilization in terms of (PD) of the constitutive tensor. In fact, eq.~\eqref{perona} always provides two negative values. The largest denotes the first failure of (PD) in a continuous decrease of $P_1$, while the latter denotes the recovery of it.

\subsubsection{Acoustic tensor and the (SE) and (E) conditions}
The acoustic tensor associated to the elasticity tensor $\mathcal{E}_{ijkl}$ is defined as 
\begin{equation}
    A_{ik} = \mathcal{E}_{ijkl} n_j n_l,
\end{equation}
where the unit vector $n_i$ is defined in the plane as
\begin{equation}
    [n_i] = [\cos \gamma, \sin \gamma],
\end{equation}
so that 
\begin{equation}
\label{acusticazzo}
    \bA(\bn) =
    \begin{bmatrix}
    \mathcal{E}_{1111} \cos^2\gamma + \mathcal{E}_{1212} \sin^2\gamma & 
    \mathcal{E}_{1221} \sin\gamma\cos\gamma \\[5mm]
    \mathcal{E}_{2112} \sin\gamma\cos\gamma & 
    \mathcal{E}_{2121} \cos^2\gamma + \mathcal{E}_{2222} \sin^2\gamma
    \end{bmatrix} ,
\end{equation}
or, more explicitly,
\begin{equation}
\label{acusticazzo2}
    \bA(\gamma) = 
    \begin{bmatrix}
    \dfrac{k_x l_x}{l_y} \cos^2\gamma + \left(\dfrac{4k_o}{l_x l_y} + \dfrac{P_2}{l_x}\right) \sin^2\gamma & 
    \dfrac{4k_o}{l_x l_y} \sin\gamma\cos\gamma \\[5mm]
    \dfrac{4k_o}{l_x l_y} \sin\gamma\cos\gamma & 
    \left(\dfrac{4k_o}{l_x l_y} + \dfrac{P_1}{l_y}\right) \cos^2\gamma + \dfrac{k_y l_y}{l_x} \sin^2\gamma
    \end{bmatrix}.
\end{equation}
The strong ellipticity (SE) condition implies that acceleration and planar waves can propagate in every direction \cite{bigoni_nonlinear_2012}. It requires that the acoustic tensor be positive definite for every unit vector $\bn$, so that for every non-null vector $\bg$
\begin{equation}
    \bg \cdot \bA(\bn) \bg > 0 \quad \Longleftrightarrow \quad \text{(SE)},  
\end{equation}
which is equivalent to the positivity of both the element $A_{11}$ of the acoustic tensor and its determinant 
\begin{multline}
\label{quartica}
\det\bA(\gamma) = 
    \underbrace{\left(\frac{4k_o}{l_xl_y}+T_{1}\right)}_{\mathcal{E}_{2121}} \frac{k_xl_x}{l_y} \cos^4 \gamma
    + \left[
    \underbrace{\frac{4k_o}{l_xl_y} \left(T_{1}+T_{2}\right) + T_{1}T_{2}}_{\text{see (PD)}} + k_xk_y
    \right] 
    \cos^2\gamma \sin^2\gamma \\
    + \underbrace{\left(\frac{4k_o}{l_xl_y}+T_{2}\right)}_{\mathcal{E}_{1212}} \frac{k_yl_y}{l_x} \sin^4 \gamma .
\end{multline}
However, assuming $l_x>0$ and $l_y>0$, and due to the requirement that (SE) holds for every $\bn$, i.e. $\gamma$, strong ellipticity reduces the positivity of the determinant of the acoustic tensor, namely,
\begin{equation}
\label{socmele}
    \det \bA(\gamma) > 0,\ \forall \gamma \in[0,2\pi] \quad \Longleftrightarrow \quad \text{(SE)}. 
\end{equation}
The ellipticity condition (E) requires that the acoustic tensor be nonsingular,
\begin{equation}
\label{ellitticazzo}
    \det \bA(\gamma) \neq 0, ~~\forall \gamma \in[0,2\pi] \quad \Longleftrightarrow \quad \text{(E)}, 
\end{equation}
so that 
\begin{quote}
\begin{center}
(PD) \quad $\Longrightarrow$ \quad (SE) \quad $\Longrightarrow$ \quad (E)   
\end{center}
\end{quote}
and the failure of (SE) in a continuous deformation path corresponds to the first failure of (E).

When ellipticity fails, a localization of incremental deformation, in other words, a discontinuity in  the gradient of incremental displacement $\nabla \bv$ can occur across a planar band in the form \cite{bigoni_nonlinear_2012} 
\begin{equation}
\label{modazzo}
    [[\nabla \bv]] = \bg_{\text{cr}} \otimes \bn_{\text{cr}},
\end{equation}
where $\bn_{\text{cr}}$ is a critical value of $\bn$ for which the vector $\bg_{\text{cr}}$ becomes an eigenvector of the acoustic tensor, associated with a null eigenvalue. The operator $[[\cdot]]$ denotes jump in the relevant argument. Therefore, a {\it shear band} occurs when $\bg_{\text{cr}} \cdot \bn_{\text{cr}}=0$, a {\it compaction band} when $\bg_{\text{cr}}$ and $\bn_{\text{cr}}$ are parallel, otherwise the band represents a {\it mixed mode}. 

\vspace{4mm} 

\noindent
Regarding the above (PD), (SE), and (E) conditions, the following details can be noticed. 
\begin{itemize}

\item Assuming $l_x>0$ and $l_y>0$, at failure of (PD) when either $T_1+T_2=0$ or $(T_1+T_2)\mathcal{H}_{1212} = -T_1T_2$ it is $\mathcal{E}_{1212} \mathcal{E}_{2121}=k_o^2/(l_xl_y)^2$, so that failure of (SE) cannot occur. Therefore, the two criteria fail simultaneously only when either $l_x=0$ or $l_y=0$. 

\item When the term multiplying $\cos^2\gamma \, \sin^2\gamma$ is positive in eq.~\eqref{quartica}, failure of (SE) occurs for $\gamma=\{0, \, \pi\}$ or $\gamma=\{\pi/2, \, 3\pi/2\}$, and the vanishing of the determinant of the acoustic tensor admits the two solutions $\mathcal{E}_{1212}=0$ or $\mathcal{E}_{2121}=0$. In these cases, the eigenvectors corresponding to the null eigenvalues of the acoustic tensor are orthogonal to $\bn$, so that a shear band forms. 

A full analytical analysis of the vanishing of the determinant of the acoustic tensor \eqref{quartica} is awkward, so that failure of the conditions $\mathcal{E}_{2121}>0$ is now examined. The component $\mathcal{E}_{2121}=\mathcal{H}_{2121}+T_1$ remains positive under tensile stress and also at small levels of compressive stress, but (i.) it may vanish and (ii.) it may become positive again with increasing compression. In particular, the vanishing of $\mathcal{E}_{2121}$ leads to the following two critical values for the force $P_1$, both corresponding to the vanishing of the determinant of the acoustic tensor
\begin{equation}
\label{perazza}
    \frac{P_1}{k_xl^0_x} = 
    -\frac{1}{2} \left(1 \pm \sqrt{1-\frac{16k_o}{k_x(l^0_x)^2}}\,\right),
\end{equation}
so that at the higher value of the critical forces, the equivalent material loses stability, while at the lower value, returns to be stable. In other words, the \lq $\pm$' sign represents the signature of restabilization.

\end{itemize}

A comparison between the (PD) condition in the case $l_yP_2>0$, eq. (\ref{perona}), and the failure of (E) in the case $\mathcal{E}_{2121}=0$, eq. (\ref{perazza}), leads to the following conclusions:

\begin{quote}
{\it In a continuous increase of compressive load, (PD) vanishes before (SE). Subsequently, with further compression, the condition (SE) may be recovered, and later also (PD). This is the essence of (i.) material instability with the failure of the exclusion condition for bifurcation occurring before the loss of ellipticity, and (ii.) subsequent restabilization, with a return inside (SE), where localization is impossible, and later inside (PD), where all bifurcations are excluded}.    
\end{quote}

The above analysis is confirmed by the examples reported in Section \ref{esempioni}, which demonstrate that for certain paths in the $P_1$--$P_2$ space, restabilization does occur,  while it remains impossible for other paths. 

It is noted in the conclusion of this section that, with the exception of the model proposed in \cite{bosi_asymptotic_2016}, all restabilizing structures introduced so far \cite{tarnai_destabilizing_1980, feodosyev_selected_1977, potier-ferry_foundations_1987, koutsogiannakis_double_2023} do not allow for the so-called \lq self-restabilization', in which the structure spontaneously restabilizes with increasing load. Rather, after the first bifurcation, the structure must be forced to remain on the trivial path to observe restabilization. This feature is retained in the equivalent continuum, because localized deformations prevail and persist after the first violation of ellipticity. In this way, the homogeneity of the state would be destroyed. Therefore, to observe restabilization in a continuum, one must postulate that the development of localized deformation bands is momentarily \lq frozen' after the first loss of ellipticity and until its recovery.

\section{A complete analysis of the unit cell as a discrete structure} \label{discrcr}

The unit cell has been analyzed in Section \ref{modellazzo} as an elementary part of an infinite grid {\it subject to a linear displacement field}, so that only 4 degrees of freedom enter the calculation. 

The purpose of this section is to provide a complete analysis of the unit cell for a general non-homogeneous deformation, revealing that: 
\begin{itemize}

\item The positive definiteness of its total potential energy leads to the same condition that defines (PD) in the equivalent solid. 

\item The exclusion of a macroscopic bifurcation for a Floquet-Bloch displacement representation {\it at small wavenumber} yields a condition coincident with the loss of (SE) in the equivalent continuum. 

\item The stability analysis in the equivalent continuum in terms of (PD) and (SE) coincides with a bifurcation analysis in the grid, with the only possible exception occurring when a micro bifurcation in the grid develops before a macro bifurcation. The latter case was never observed in any of the examples referred to the structures considered in the present article.

\end{itemize}

The fact that micro bifurcations may occur while the response of the equivalent material remains unaffected is demonstrated in \cite{geymonat_homogenization_1993,triantafyllidis_comparison_1993,lopez-pamies_overall_2006}. The (PD) condition was considered in the context of homogenization until now only in \cite{bordiga_dynamics_2021}.

\begin{figure}[hbt!]
    \centering
    \includegraphics[width=0.5\linewidth]{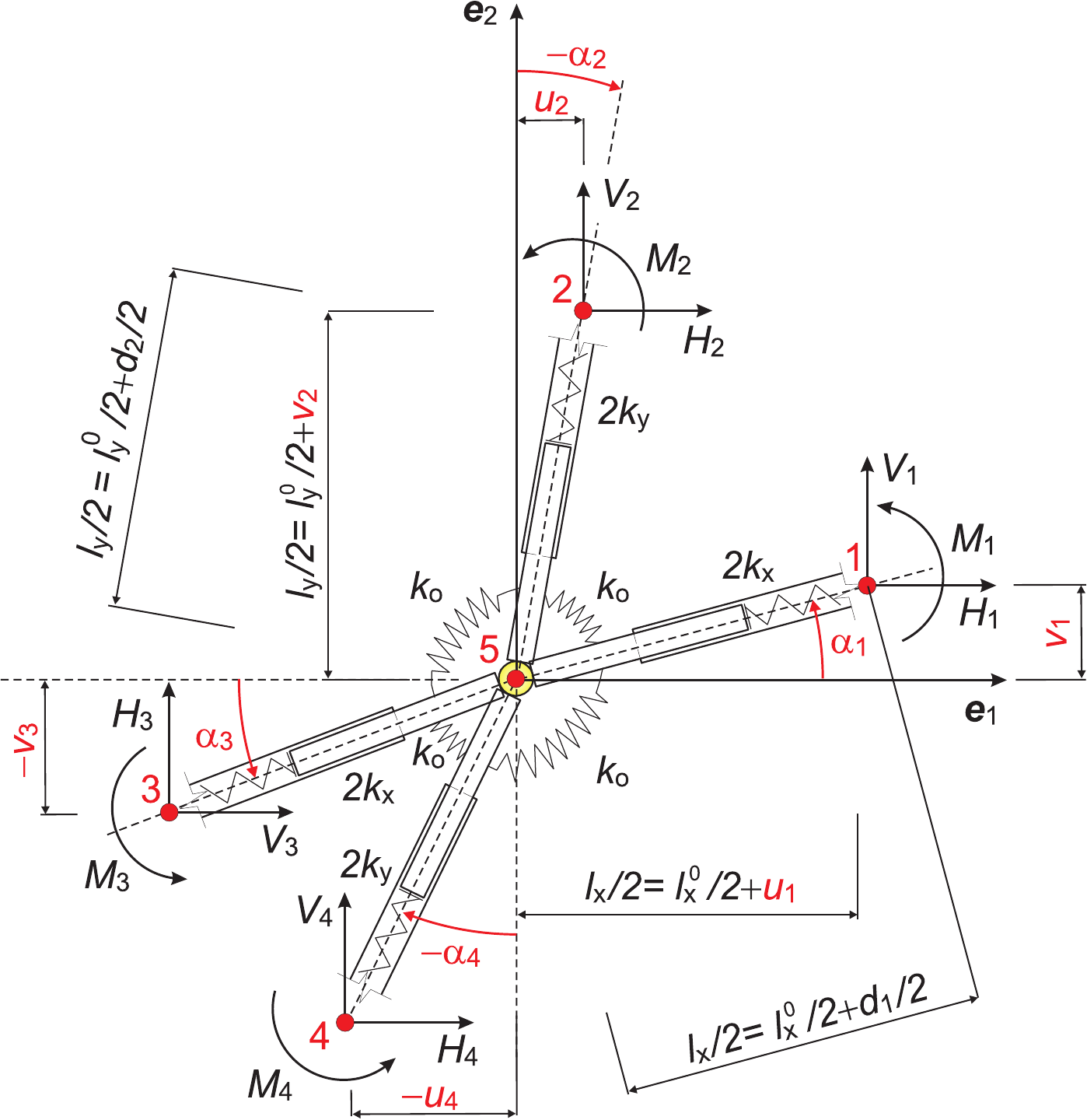}
    \caption{The deformation of the unit cell when subject to generic dead forces $H_i$, $V_i$ and $M_i$, $i=1,...,4$ applied on the bar ends. Components $u_i$ and $v_i$ denote the horizontal and vertical displacements of the nodes $i=1,...,5$. The node \#5 is central, and its displacement is not shown in the figure; the longitudinal springs elongate of an amount $d_i$, $i=1,...,4$, so that only one half is pertinent to the half bar. Note that the bars are flexurally rigid, so that their rotations $\alpha_i$ are related to the end displacements.}
    \label{maglia_rigida_5}
\end{figure}

The unit cell is a structure characterized by 10 degrees of freedom, the displacements of the five nodes (the central junction, labelled \#5, plus the four boundary nodes, labelled from \#1 to \#4)
\begin{equation}
    \bq = \{u_1,v_1,u_2,v_2,u_3,v_3,u_4,v_4,u_5,v_5\}^T,
\end{equation}
where $u$ and $v$ denote horizontal and vertical components, respectively, as shown in Fig.~\ref{maglia_rigida_5}. 
%

Noting that only one-half of the axial springs with stiffnesses $k_x$ and $k_y$ pertains to the unit cell (from which the factor $1/4$ enters the equation below), the total potential energy in a deformed configuration is given by
\begin{multline}
\label{eq:TPE}
    W(\bq) = 
    \frac{k_o}{2} (\alpha_1-\alpha_2)^2 + \frac{k_o}{2} (\alpha_2-\alpha_3)^2 
    + \frac{k_o}{2} (\alpha_3-\alpha_4)^2 + \frac{k_o}{2} (\alpha_4-\alpha_1)^2 
    + \frac{1}{4} k_xd_1^2 + \frac{1}{4} k_yd_2^2 \\
    + \frac{1}{4} k_xd_3^2 + \frac{1}{4} k_yd_4^2
    - H_1\frac{l^0_x}{2} \left[\left(1+\frac{d_1}{l^0_x}\right)\cos\alpha_1 - 1\right] 
    - V_1\frac{l^0_x}{2} \left(1+\frac{d_1}{l^0_x}\right)\sin\alpha_1 - M_1 \alpha_1 \\ 
    - V_2\frac{l^0_y}{2} \left[\left(1+\frac{d_2}{l^0_y}\right)\cos\alpha_2 - 1\right] 
    + H_2\frac{l^0_y}{2} \left(1+\frac{d_2}{l^0_y}\right)\sin\alpha_2  - M_2 \alpha_2
    + H_3\frac{l^0_x}{2} \left[\left(1+\frac{d_3}{l^0_x}\right)\cos\alpha_3 - 1\right] \\ 
    + V_3\frac{l^0_x}{2} \left(1+\frac{d_3}{l^0_x}\right)\sin\alpha_3 - M_3 \alpha_3 
    + V_4\frac{l^0_y}{2} \left[\left(1+\frac{d_4}{l^0_y}\right)\cos\alpha_4 - 1\right] 
    - H_4\frac{l^0_y}{2} \left(1+\frac{d_4}{l^0_y}\right)\sin\alpha_4 - M_4 \alpha_4,
\end{multline}
where the variation in the length of the four bars is
\begin{equation}
\label{eq:elongations}
\begin{aligned}
    d_1 = 2 \sqrt{\left(\frac{l_x^0}{2} + u_1 - u_5\right)^2 + (v_1 - v_5)^2} - l_x^0, \quad
    d_2 = 2 \sqrt{\left(\frac{l_y^0}{2} + v_2 - v_5\right)^2 + (u_2 - u_5)^2} - l_y^0, \\
    d_3 = 2 \sqrt{\left(\frac{l_x^0}{2} + u_5 - u_3\right)^2 + (v_3 - v_5)^2} - l_x^0, \quad
    d_4 = 2 \sqrt{\left(\frac{l_y^0}{2} + v_5 - v_4\right)^2 + (u_4 - u_5)^2} - l_y^0, 
\end{aligned}
\end{equation}
while the inclination angles of the bars can be written as
\begin{equation}
\label{eq:rotations}
\begin{aligned}
    \alpha_1 = \arctan \frac{v_1 - v_5}{l_x^0/2 + u_1 - u_5}, \quad
    \alpha_2 = \arctan \frac{u_5 - u_2}{l_y^0/2 + v_2 - v_5}, \\
    \alpha_3 = \arctan \frac{v_5 - v_3}{l_x^0/2 + u_5 - u_3}, \quad
    \alpha_4 = \arctan \frac{u_4 - u_5}{l_y^0/2 + v_5 - v_4}. 
\end{aligned}
\end{equation}
Eqs.~\eqref{eq:elongations} and \eqref{eq:rotations} show that the 10 nodal displacement components determine the configuration of the bars, as their inclination does not play a role due to the bars' rigidity under bending.

\subsection{A local exclusion condition for bifurcation: Positive definiteness of the reduced stiffness matrix of the unit cell}
\label{reticolazzo}

In a similar vein to the definition of the (PD) condition in a continuum, the aim of this section is to find an exclusion condition for bifurcation,  which may occur in any grid of axially deformable but flexurally rigid elements (the proof-of-concept model, Fig.~\ref{maglia_rigida_2}) forming any domain, and subject to arbitrary boundary conditions. The only assumed restriction is that the prestress in the unit cell is only axial, so that shear forces and bending moments are not permitted before incremental deformation. In contrast, the boundary conditions and the applied internal forces must be compatible with this condition. 

For any initial configuration which is biaxially deformed, so that the bars remain aligned parallel to their initial directions, $\alpha_1=\alpha_2=\alpha_3=\alpha_4=0$ and $Q_1=Q_2=0$, but are elongated or shortened due to the application of axial pre-loads, $P_1 \neq P_2 \neq 0$, the following displacement vector characterizes the preloaded configuration:
\begin{equation}
    \bq_0 = \{u_1^0,0,0,v_2^0,-u_1^0,0,0,-v_2^0,0,0\}^T.
\end{equation}
The stability of the above-preloaded configuration can be assessed by analysing the incremental stiffness matrix of the unit cell. For this purpose, the total potential energy, eq.~\eqref{eq:TPE}, of the unit cell is expanded asymptotically around the preloaded configuration $\bq_0$ as follows:
\begin{equation}
    W(\bq_0 + \delta{\bq}) = W(\bq_0) + \left.\frac{\partial W}{\partial \bq}\right|_{\bq_0} \cdot \delta{\bq} + \frac{1}{2} \delta{\bq} \cdot \left.\frac{\partial^2 W}{\partial \bq \partial \bq}\right|_{\bq_0} \delta{\bq} + \text{h.o.t}
\end{equation}
In the asymptotic expansion, the constant term $W(\bq_0)$ is irrelevant, because the total potential energy is defined up to a constant term. Instead, the linear term must be set equal to zero for any increment $\delta{\bq}$, to enforce the equilibrium of the preloaded configuration
\begin{equation}
    \left.\frac{\partial W}{\partial \bq}\right|_{\bq_0} \cdot \delta{\bq} = 0 \quad \forall \delta{\bq},
\end{equation}
which provides the pre-load within the longitudinal springs and, again, the condition of vanishing $Q_1$ and $Q_2$, two components of a kind of generalized shear
\begin{equation}
\begin{aligned}
    & P_1 = H_1 = -H_2 =2 k_x u_1^0, \quad P_2 = V_2 = -V_4 = 2 k_y v_2^0, \\[3mm] 
    & Q_1 = V_1 + \frac{M_1}{\frac{l_x^0}{2} + u_1^0} = 
    - V_3 + \frac{M_3}{\frac{l_x^0}{2} + u_1^0} = 0, \\[3mm] 
    & Q_2 = H_2 + \frac{M_2}{\frac{l_y^0}{2} + v_2^0} = 
    -H_4 + \frac{M_4}{\frac{l_y^0}{2} + v_2^0} = 0.
\end{aligned}
\end{equation}
Therefore, up to the quadratic term, the total potential energy of the preloaded configuration reduces to 
\begin{equation}
    W(\bq_0 + \delta{\bq}) = \frac{1}{2} \delta{\bq} \cdot \bK_0 \delta{\bq},
\end{equation}
where
\begin{equation}
    \bK_0(P_1,P_2) = \left.\frac{\partial^2 W}{\partial \bq \partial \bq}\right|_{\bq_0},
\end{equation}
is the incremental stiffness matrix of the unit cell, which depends on the pre-load $\bP=\{P_1,P_2\}$ and can be written as 
\begin{multline}
\label{eq:K0}
    \bK_0(P_1,P_2) = \\
    \begin{bmatrix}
    2 k_x & 0 & 0 & 0 & 0 & 0 & 0 & 0 & -2 k_x & 0 \\
    0 & K_{22} & K_{23} & 0 & 0 & 0 & -K_{23} & 0 & 0 & -K_{22} \\
    0 & K_{23} & K_{33} & 0 & 0 & -K_{23} & 0 & 0 & -K_{33} & 0 \\
    0 & 0 & 0 & 2 k_y & 0 & 0 & 0 & 0 & 0 & -2 k_y \\
    0 & 0 & 0 & 0 & 2 k_x & 0 & 0 & 0 & -2 k_x & 0 \\
    0 & 0 & -K_{23} & 0 & 0 & K_{22} & K_{23} & 0 & 0 & -K_{22} \\
    0 & -K_{23} & 0 & 0 & 0 & K_{23} & K_{33} & 0 & -K_{33} & 0 \\
    0 & 0 & 0 & 0 & 0 & 0 & 0 & 2 k_y & 0 & -2 k_y \\
    -2 k_x & 0 & -K_{33} & 0 & -2 k_x & 0 & -K_{33} & 0 & 2K_{33} + 4k_x & 0 \\
    0 & -K_{22} & 0 & -2 k_y & 0 & -K_{22} & 0 & -2 k_y & 0 & 2K_{22} + 4k_y \\
    \end{bmatrix},
\end{multline}
where
\begin{equation}
\begin{aligned}
    K_{22} &= \frac{2 k_x (4 k_o k_x+P_1 (k_x l_x^0+P_1))}{(k_x l_x^0+P_1)^2} = \frac{8k_o+2l_xP_1}{l_x^2}, \\[3mm]
    K_{33} &= \frac{2 k_y (4 k_o k_y+P_2 (k_y l_y^0+P_2))}{(k_y l_y^0+P_2)^2} = \frac{8k_o+2l_yP_2}{l_y^2}, \\[3mm]
    K_{23} &= \frac{4 k_o k_x k_y}{(k_x l_x^0+P_1) (k_y l_y^0+P_2)} = \frac{4k_0}{l_x l_y}.
\end{aligned}
\end{equation}
It is worth noting that the incremental stiffness matrix $\bK_0$ is always singular, as two linearly independent rigid-body translations are possible for the two-dimensional unit cell. Thus, the dimension of the nullspace of $\bK_0$ equals two. A sufficient condition for the stability of the lattice can be obtained following these two operations: (i) remove the two rigid-body translations by fixing the central node, $u_5 = v_5 = 0$, and consequently reduce the incremental stiffness matrix by eliminating its two last rows and columns; (ii) impose the positive definiteness of the reduced $8 \times 8$ stiffness matrix. 

The eight eigenvalues of the reduced matrix are: 
\begin{equation}
    2 k_x \text{ (double)}, \quad 2 k_y \text{ (double)}, \quad K_{22}, \quad K_{33}, \quad
    \frac{1}{2} \left(K_{22}+K_{33} \pm \sqrt{(K_{22}-K_{33})^2 + 16 K_{23}^2}\right), 
\end{equation}
and thus, a sufficient condition for the  uniqueness of the incremental response 
of the lattice is
\begin{equation}
\label{pidone}
    K_{22}+K_{33} - \sqrt{(K_{22}-K_{33})^2 + 16 K_{23}^2} > 0,
\end{equation}
{\it which coincides with the counterpart in the equivalent solid, namely, the (PD) condition, eq.~\eqref{pd} or eq.~\eqref{orchite}. 
}

Condition \eqref{pidone} excludes any bifurcation for all combinations of cells satisfying it. This holds even if the grid occupies a finite domain and for every combination of mixed boundary conditions applied at its edges.

It should finally be noted that the coincidence between the (PD) condition in the equivalent solid and the positive definiteness of the 8 $\times$ 8 reduction of the stiffness matrix $\bK_0$ holds for the proof-of-principle structure shown in Fig.~\ref{maglia_rigida_2}. More in general, when the flexurally rigid bars are replaced by flexible rods, the coincidence does not hold, as will be shown with a counterexample in Section~\ref{isolotti}, Fig.~\ref{fig:stability_domains_cubic_deform_200}.

\subsection{An exclusion condition for bifurcation of an infinite, periodic distribution of cells: Positive definiteness  of the matrix governing Bloch waves}

When a periodic ensemble of identical cells is considered, distributed in a finite (or infinite) region of the plane, and the cells are all uniformly prestressed and subjected to prescribed displacements on the entire boundary of the region (possibly at infinity), an equivalent of the van Hove theorem for a solid \cite{bigoni_nonlinear_2012}, but now for a discrete grid, can be proven (the proof will be given elsewhere). 
This theorem leads to the conclusion that the positive definiteness of the matrix governing Bloch wave propagation is sufficient to enforce the uniqueness of the incremental response. Therefore, as the failure of (SE) in an infinite and uniformly deformed elastic body is the condition relevant to bifurcation and strain localization, the analogous condition for the grid is played by the positive definiteness of the stiffness matrix $\bK_0^*$, which will be derived below. However, while in the continuous medium, the loss of (SE) is synonymous with strain localization, the loss of positive definiteness of $\bK_0^*$ may occur in both micro and macro modes, and only the latter corresponds to loss of (SE) in the equivalent material \cite{michel_microscopic_2007,michel_microscopic_2010,santisi_davila_localization_2016}. In fact, it is proven in this section that the condition for the loss of positive definiteness of $\bK_0^*$ coincides, at the vanishing wavenumber limit (i.e., infinite wavelength limit), with the (SE) condition for the equivalent material, eq.~\eqref{quartica}.

The incremental equilibrium equations for the unit cell in the preloaded configuration, subjected to incremental forces on its boundary nodes, are considered
\begin{equation}
\label{eq:equilibrium}
    \bK_0(P_1,P_2) \dot{\bq} = \dot{\bef},
\end{equation}
where $\dot{\bq}$ and $\dot{\bef}$ denote the incremental displacements and incremental forces, respectively,
\begin{equation}
\label{eq:vectors}
    \dot{\bq} = \{\dot{u}_1,\dot{v}_1,\dot{u}_2,\dot{v}_2,\dot{u}_3,\dot{v}_3,\dot{u}_4,\dot{v}_4,\dot{u}_5,\dot{v}_5\}^T, \quad 
    \dot{\bef} = \{\dot{H}_1,\dot{V}_1,\dot{H}_2,\dot{V}_2,\dot{H}_3,\dot{V}_3,\dot{H}_4,\dot{V}_4,\dot{H}_5,\dot{V}_5\}^T.   
\end{equation}
In the absence of body forces, the nodes inside the unit cell are not loaded from external forces, $\dot{H}_5=\dot{V}_5=0$. Moreover, the forces acting on the boundary nodes are \emph{internal to the lattice}, i.e. they are exerted on the unit cell by the neighbouring cells.

A quasi-periodic non-trivial solution of eqs.~\eqref{eq:equilibrium} is assumed according to Bloch-wave theory. This solution is governed by the following Floquet-Bloch conditions on the boundary nodes of the unit cell
\begin{equation}
\label{eq:conditions}
\begin{aligned}
    \dot{u}_1 &= \dot{u}_3 e^{i k_1 l_x}, & \dot{v}_1 &= \dot{v}_3 e^{i k_1 l_x}, & \dot{\alpha}_1 &= \dot{\alpha}_3 e^{i k_1 l_x}, \\
    \dot{u}_2 &= \dot{u}_4 e^{i k_2 l_y}, & \dot{v}_2 &= \dot{v}_4 e^{i k_2 l_y}, & \dot{\alpha}_2 &= \dot{\alpha}_4 e^{i k_2 l_y}, \\
\end{aligned}
\end{equation}
where $\bk = \{k_1,k_2\}$ is the Bloch vector and $l_x=l_x^0+P_1/k_x$, $l_y=l_y^0+P_2/k_y$ are the current lengths of the bars. The four incremental rotations $\dot{\alpha}_j$, $j=1,\cdots,4$, are related to the node incremental displacements through eqs.~\eqref{eq:rotations}, which in a linearised setting (valid for small-amplitude modes), become
\begin{equation}
    \dot{\alpha}_1 = \frac{2}{l_x} (\dot{v}_1 - \dot{v}_5), \quad 
    \dot{\alpha}_2 = -\frac{2}{l_y} (\dot{u}_2 - \dot{u}_5), \quad 
    \dot{\alpha}_3 = -\frac{2}{l_x} (\dot{v}_3 - \dot{v}_5), \quad 
    \dot{\alpha}_4 = \frac{2}{l_y} (\dot{u}_4 - \dot{u}_5).
\end{equation}
Consequently, the Floquet-Bloch conditions can be rewritten in terms of incremental displacements only (involving now also the central node),
\begin{equation}
\begin{aligned}
    \dot{u}_1 &= \dot{u}_3 e^{i k_1 l_x}, & \dot{v}_1 &= \dot{v}_5 \frac{1+e^{i k_1 l_x}}{2}, & \dot{v}_3 &= \dot{v}_5 \frac{1+e^{-i k_1 l_x}}{2}, \\
    \dot{u}_2 &= \dot{u}_5 \frac{1+e^{i k_2 l_y}}{2}, & \dot{v}_2 &= \dot{v}_4 e^{i k_2 l_y}, & \dot{u}_4 &= \dot{u}_5 \frac{1+e^{i k_2 l_y}}{2}, \\
\end{aligned}
\end{equation}
so that the independent degrees of freedom reduce to four: $\dot{u}_5$, $\dot{v}_5$, $\dot{u}_3$ and $\dot{v}_4$. These conditions can be conveniently expressed in the matrix form
\begin{equation}
\label{eq:bloch}
    \dot{\bq} = 
    \begin{Bmatrix}
        \dot{u}_1 \\ \dot{v}_1 \\ \dot{u}_2 \\ \dot{v}_2 \\ \dot{u}_3 \\ \dot{v}_3 \\ \dot{u}_4 \\ \dot{v}_4 \\ \dot{u}_5 \\ \dot{v}_5
    \end{Bmatrix}
    = 
    \begin{bmatrix}
        0 & 0 & z_1 & 0 \\
        0 & (1+z_1)/2 & 0 & 0 \\
        (1+z_2)/2 & 0 & 0 & 0 \\
        0 & 0 & 0 & z_2 \\
        0 & 0 & 1 & 0 \\
        0 & (1+\bar{z}_1)/2 & 0 & 0 \\
        (1+\bar{z}_2)/2 & 0 & 0 & 0 \\
        0 & 0 & 0 & 1 \\
        1 & 0 & 0 & 0 \\
        0 & 1 & 0 & 0 \\
    \end{bmatrix}
    \begin{Bmatrix}
        \dot{u}_5 \\ \dot{v}_5 \\ \dot{u}_3 \\ \dot{v}_4
    \end{Bmatrix},
\end{equation}
concisely rewritten as
\begin{equation}
\label{eq:bloch2}
    \dot{\bq} = \bZ(k_1,k_2) \dot{\bq}^*,
\end{equation}
where $\bZ(k_1,k_2)$ and $\dot{\bq}^*$ are defined accordingly to eq.~\eqref{eq:bloch} and $z_1 = e^{i k_1 l_x}$, $z_2 = e^{i k_2 l_y}$. A substitution of eq.~\eqref{eq:bloch} into eq.~\eqref{eq:equilibrium} gives
\begin{equation}
\label{pisello}
    \bK_0(P_1,P_2) \bZ(k_1,k_2) \dot{\bq}^* = \dot{\bef}.
\end{equation}
A left multiplication of eq.~\eqref{pisello} by $\bZ(k_1,k_2)^{\mathsf{H}}$, where the superscript ${\mathsf{H}}$ denotes the complex conjugate transpose operation, leads to the reduced system
\begin{equation}
    \bZ(k_1,k_2)^{\mathsf{H}} \bK_0(P_1,P_2) \bZ(k_1,k_2) \dot{\bq}^* = \dot{\bef}^*,
\end{equation}
where
\begin{equation}
    \dot{\bef}^* = \bZ(k_1,k_2)^{\mathsf{H}} \dot{\bef} = 
    \begin{Bmatrix}
        \dfrac{\dot{H}_2 + z_2 \dot{H}_4}{2} + \dfrac{\dot{H}_4 + \bar{z}_2 \dot{H}_2}{2} \\
        \dfrac{\dot{V}_1 + z_1 \dot{V}_3}{2} + \dfrac{\dot{V}_3 + \bar{z}_1 \dot{V}_1}{2} \\
        \dot{H}_3 + \bar{z}_1 \dot{H}_1 \\
        \dot{V}_4 + \bar{z}_2 \dot{V}_2
    \end{Bmatrix}.
\end{equation}
Note that if the boundary displacements satisfy the quasi-periodicity conditions \eqref{eq:conditions}, then the boundary forces will also satisfy the same conditions but with the opposite sign since the boundary nodes must be in equilibrium,
\begin{equation}
\label{eq:forces}
\begin{aligned}
    \dot{H}_1 &= -\dot{H}_3 e^{i k_1 l_x}, & \dot{V}_1 &= -\dot{V}_3 e^{i k_1 l_x}, \\
    \dot{H}_2 &= -\dot{H}_4 e^{i k_2 l_y}, & \dot{V}_2 &= -\dot{V}_4 e^{i k_2 l_y}, 
\end{aligned}
\end{equation}
so that 
\begin{equation}
\begin{aligned}
    \dot{H}_2 + z_2 \dot{H}_4 &= \dot{H}_4 +\bar{z}_2 \dot{H}_2 = 0, \\
    \dot{V}_1 + z_1 \dot{V}_3 &= \dot{V}_3 +\bar{z}_1 \dot{V}_1 = 0, \\
    \dot{H}_3 + \bar{z}_1 \dot{H}_1 &= 0, \\
    \dot{V}_4 + \bar{z}_2 \dot{V}_2 &= 0,
\end{aligned}
\end{equation}
which implies $\dot{\bef}^* = \bzero$. 

Therefore, the following homogeneous system is obtained
\begin{equation}
\label{eq:homogeneous}
    \bK_0^*(k_1,k_2,P_1,P_2) \dot{\bq}^* = \bzero,
\end{equation}
where the matrix 
\begin{equation}
\label{poldino}
    \bK_0^*(k_1,k_2,P_1,P_2) = \bZ(k_1,k_2)^{\mathsf{H}} \bK_0(P_1,P_2) \bZ(k_1,k_2) 
\end{equation}
is
\begin{multline}
\label{eq:reduced}
    \bK_0^* 
    = \begin{bmatrix}
        K^*_{11} & K^*_{12} & -2(1+e^{i k_1l_x}) k_x & 0 \\
        K^*_{21} & K^*_{22} & 0 & -2(1+e^{i k_2l_y}) k_y \\
        -2(1+e^{-i k_1l_x}) k_x & 0 & 4 k_x & 0 \\
        0 & -2(1+e^{-i k_2l_y}) k_y & 0 & 4 k_y
    \end{bmatrix}, 
\end{multline}
in which the elements 
\begin{equation}
\begin{aligned}
    K^*_{11} &= \frac{1}{l_y^2} [8k_o + 2l_y(2k_xl_y+P_2) - 2(4k_o+l_yP_2)\cos k_2 l_y], \\
    K^*_{22} &= \frac{1}{l_x^2} [8k_o + 2l_x(2k_yl_x+P_1) - 2(4k_o+l_xP_1)\cos k_1 l_x], \\
    K^*_{12} &= K^*_{21} = \frac{4k_o}{l_x l_y}\sin k_1l_x \sin k_2l_y
\end{aligned}
\end{equation}
\lq condense' the dependence on $P_1$ and $P_2$.

A non-trivial solution of eq.~\eqref{eq:homogeneous} and, hence, a bifurcation of the lattice, becomes possible when a Bloch vector $\bk = \{k_1,k_2\}$ and a preload $\bP=\{P_1,P_2\}$ exist such that 
\begin{multline}
\label{eq:det}
    \det \bK_0^*(k_1,k_2,P_1,P_2) = 
    16 k_x k_y \Big[ 
    (K^*_{11} - 2k_x) (K^*_{22} - 2k_y) - (K^*_{12})^2 \\
    - 2 k_x (K^*_{22} - 2k_y) \cos k_1 l_x  
    - 2 k_y (K^*_{11} - 2k_x) \cos k_2 l_y
    + 4 k_x k_y \cos k_1 l_x \cos k_2 l_y
    \Big] = 
    0.
\end{multline}
Note that the matrix $\bK_0^*(\bk,\bP)$ is Hermitian. Thus, the determinant \eqref{eq:det} is always real. Moreover, the periodicity of $\bZ(\bk)$ implies that this determinant is periodic in the $\bk$-space with period $[0, 2\pi]\times[0, 2\pi]$ in the basis $\{\bb_1,\bb_2\}$, reciprocal to the lattice direct basis $\{\ba_1,\ba_2\}$, so that $\bb_i \cdot \ba_j = \delta_{ij}$. 
It is clear from eq.~\eqref{eq:det} that the bifurcation modes can exhibit different wavelengths $2\pi/k$, where $k = |\bk| = \sqrt{k_1^2+k_2^2}$. When the wavelength becomes infinite, a \emph{global} or \emph{macro} bifurcation occurs; otherwise, the bifurcation is called \emph{microscopic}. 

Therefore, a macroscopic bifurcation of the lattice can be identified by assuming the wave vector in the form $\bk = k \bn$, where $\bn = \{n_1,n_2\}$ is a unit vector defining the wave direction, and then considering the leading-order term of eq.~\eqref{eq:det} in an expansion as $k \to 0$. This gives
\begin{multline}
\label{poldone}
    16 k_x k_y l_x^2 l_y^2 
    \left\{
    \left(\frac{4k_o}{l_xl_y} + \frac{P_1}{l_y}\right) \frac{k_xl_x}{l_y} n_1^4 
    + \left[\frac{4k_o}{l_xl_y}\left(\frac{P_1}{l_y} + \frac{P_2}{l_x}\right) + \frac{P_1P_2}{l_xl_y} + k_xk_y\right] n_1^2 n_2^2 \right. \\
    \left. + \left(\frac{4k_o}{l_xl_y} + \frac{P_2}{l_x}\right) \frac{k_yl_y}{l_x} n_2^4
    \right\} = 0,
\end{multline}
which coincides with the failure of ellipticity for the equivalent elastic solid, namely, the vanishing of the determinant of the acoustic tensor, eq.~\eqref{quartica}, except for the presence of the strictly positive factor $16 k_x k_y l_x^2 l_y^2$. This shows that, for the proof-of-principle structure shown in Fig.~\ref{maglia_rigida_2}, {\it the macroscopic bifurcation of the lattice is equivalent to the loss of ellipticity of the effective continuum}. Conversely, microscopic bifurcations of the lattice remain undetected in the effective continuum. In any case, in all our examples related to the proof-of-principle structure, macro bifurcations have been found to be the first manifestation of instability in the grid. 

It is finally remarked that stability in terms of Floquet-Bloch wave propagation in the lattice requires that all the eigenvalues of the matrix $\bK_0^*$, eq.~\eqref{eq:reduced}, be strictly positive. 

The positive definiteness of $\bK_0^*$ plays an important role, because it is used to demonstrate that the restabilization occurring in the equivalent elastic solid is a \lq true' restabilization, holding also for the underlying lattice. This is a crucial point distinguishing the \lq true' restabilization found in the present article from the mere restabilization of the effective medium, taking place while the underlying elastic grid remains unstable, as found in \cite{bordiga_tensile_2022} and also below, see Fig.~\ref{fig:stability_domains_cubic_deform_200} in Section~\ref{isolotti}.

\section{A homogenization based on the calculus of variations and notions of Gamma-convergence}
\label{gammastokk}

The homogenization scheme presented in Section \ref{modellazzo} is based on purely mechanical considerations. We have privileged this formulation because shear bands and restabilization are obtained as genuine mechanical phenomena and are not hindered by mathematical technicalities. 
However, the reader might question whether the same results on homogenization could be obtained through a more formal analysis based on the calculus of variations and notions of Gamma-convergence (see, for instance, \cite{ariza_homogenization_2024}, where grids of elastic rods are addressed, but in the absence of prestress). The answer to this question is positive, so that the purpose of the present Section is to show that:

\begin{quote}
    Formal homogenization based on variational calculus leads exactly to the elasticity tensor \eqref{paola_gatti}. 
\end{quote}

The presentation is limited to the homogenization algorithm based on variational calculus and $\Gamma$-convergence, while its mathematical foundations are not reviewed for conciseness; for this aspect, the interested reader is addressed to \cite{ariza_homogenization_2024}. 

\begin{itemize}

    \item The homogenization procedure concerns the limit where the lattice cell size is much smaller than the domain size, also known as the {\it continuum limit}. The limiting energy density of the composite is independent of the domain it occupies. Therefore, the elastic energy density of the effective continuum can be conveniently determined by considering a lattice of infinite extent. This lattice is generated by tessellating the unit cell shown in Fig.~\ref{maglia_rigida_2} along the two orthogonal directions $\{\be_1,\be_2\}$, which define the direct basis of the lattice. Each unit cell is uniquely identified by a pair of indices $(m, n)$, and denoted as $\mC_{m,n}$. Consequently, the infinite lattice domain $\mB$ can be expressed as the union of all unit cells indexed by the set $\fZ^2$, such that
    \begin{equation}
    \label{dominio}
        \mB = \bigcup_{(m,n) \in \fZ^2} \mC_{m,n}.
    \end{equation}
    As a consequence, the elastic energy of the infinite grid is 
    \begin{equation}
        W(\bq) = \sum_{(m,n) \in \fZ^2} \frac{1}{2} \bq_{m,n} \cdot \bK_0 \bq_{m,n}
    \end{equation}
    where $\bK_0$ is the incremental stiffness matrix of the unit cell, eq.~\eqref{eq:K0}, and $\bq_{m,n}$ is the incremental displacement vector of the unit cell $\mC_{m,n}$, eq.~\eqref{eq:vectors}$_1$. 

    \item The discrete time Fourier transform (DTFT) is defined for the lattice as
    \begin{equation}
        \widehat{\bq}(\bk) = V \sum_{(m,n) \in \fZ^2} \bq_{m,n} e^{-i(k_1 l_x m + k_2 l_y n)},
    \end{equation}
    where $V = l_x l_y$ is the current volume of the unit cell $\mC_{m,n}$. 
    The inverse transform is 
    \begin{equation}
        \bq_{m,n} = \frac{1}{4 \pi^2} \int_{\mC^*} \bq(\bk) e^{i(k_1 l_x m + k_2 l_y n)} d\bk , 
    \end{equation}
    where $\mC^* = \left[-\frac{\pi}{l_x},\frac{\pi}{l_x}\right] \times \left[-\frac{\pi}{l_y},\frac{\pi}{l_y}\right]$ is the reciprocal unit cell.
    An application of the DTFT to the $10 \times 1$ displacement vector $\bq_{m,n}$ leads to 
    \begin{equation}
    \label{eq:dtft}
        \widehat{\bq}(k_1,k_2) = \bZ(k_1,k_2) \widehat{\bq}^*(k_1,k_2),
    \end{equation}
    where $\bZ(k_1,k_2)$ is the $10 \times 4$ matrix given in eq.~\eqref{eq:bloch} and $\widehat{\bq}^*(\bk)$ is the $4 \times 1$ reduced displacement vector
    \begin{equation}
        \widehat{\bq}^*(\bk) = \{\widehat{u}_5(\bk), \widehat{v}_5(\bk), \widehat{u}_3(\bk), \widehat{v}_4(\bk)\}.
    \end{equation}  

    \item The Parseval's theorem allows to represent the elastic energy as 
    \begin{equation}
        W = \frac{1}{4\pi^2 V} \int_{\mC^*} \frac{1}{2} \widehat{\bq}(\bk) \cdot \bK_0 \overline{\widehat{\bq}(\bk)}\, d\bk, 
    \end{equation}
    where a bar over a symbol denotes the complex conjugate, so that, a substitution of eq.~\eqref{eq:dtft} yields 
    \begin{equation}
        W = \frac{1}{4\pi^2 V} \int_{\mC^*} \frac{1}{2} \widehat{\bq}(\bk) \cdot \Big[\bZ(\bk)^H \bK_0 \bZ(\bk)\Big] \overline{\widehat{\bq}(\bk)}\, d\bk,
    \end{equation}
    where the superscript $^H$ denotes the conjugate transpose operator, while 
    \begin{equation}
        \bK_0^*(\bk) = \bZ(\bk)^H \bK_0 \bZ(\bk) , 
    \end{equation}
    is the matrix given in eq.~\eqref{eq:reduced}, referred to as the \lq dynamical matrix' in \cite{ariza_homogenization_2024}.
    
    \item Scaling is introduced by considering a sequence of lattices spanning increasingly larger self-similar domains $\Omega/\varepsilon$, where $\varepsilon > 0$ is a small scaling factor. The sequence of scaling energies takes the form 
    \begin{equation}
        W_{\varepsilon} = \frac{1}{4\pi^2 V} \int_{\mC^*_{\varepsilon}} \frac{1}{2} \widehat{\bq}(\bk) \cdot \bK_{0,\varepsilon}^*(\bk) \overline{\widehat{\bq}(\bk)}\, d\bk, \qquad \mC^*_{\varepsilon} = \left[-\frac{\pi}{\varepsilon l_x},\frac{\pi}{\varepsilon l_x}\right] \times \left[-\frac{\pi}{\varepsilon l_y},\frac{\pi}{\varepsilon l_y}\right] , 
    \end{equation}
    where the stiffness matrix scales as 
    \begin{equation}
        \bK_{0,\varepsilon}^*(\bk) = \varepsilon^{-2} \bK_{0}^*(\varepsilon \bk).
    \end{equation}
    
    \item Assume that the lattice is acted upon by macroscopic distributed forces defined by the function
    \begin{equation}
    \label{forzedelcaz}
        \bb_0(\bx) = (b_{0x}(\bx), b_{0y}(\bx)),
    \end{equation}
    then, the equilibrium equations in the transformed space become 
    \begin{equation}
        \bK_{0,\varepsilon}^*(\bk)\, \widehat{\bq}^*_\varepsilon(\bk) = \widehat{\bb}(\bk), \qquad \bk \in \mC^*_{\varepsilon},
    \end{equation}
    where 
    \begin{equation}
        \widehat{\bb}(\bk) = \mL\, \widehat{\bb}_0(\bk) ,
    \end{equation}
    in which $\widehat{\bb}_0(\bk)$ is the ordinary Fourier transform of $\bb_0(\bx)$ and the operator $\mL$ localizes the continuum forces to each of the lattice joints, so that it has the form
    \begin{equation}
        \mL =
        \begin{bmatrix}
            1/2 & 0 \\
            0 & 1/2 \\
            1/2 & 0 \\
            0 & 1/2 
        \end{bmatrix}.
    \end{equation}
    Note that our choice of macroscopic distributed forces, eq.~(\ref{forzedelcaz}), does not include any couple, as this would not be admissible for a Cauchy elastic effective material. Therefore, our equivalent solid turns out to be a Cauchy material, where micropolar effects are excluded. 

    \item The \lq effective dynamical matrix' in \cite{ariza_homogenization_2024} is obtained as the continuum limit 
    \begin{equation}
        \bK_{0}^{\text{EFF}}(\bk) = \left( \lim_{\varepsilon \to 0} \mL^T \left(\frac{1}{V} \bK_{0,\varepsilon}^*(\bk) \right)^{-1} \mL \right)^{-1},
    \end{equation}
    which evaluates explicitly as
    \begin{equation}
        \bK_{0}^{\text{EFF}}(\bk) = 
        \begin{bmatrix}
        \dfrac{k_x l_x}{l_y} k_1^2 + \left(\dfrac{4k_o}{l_x l_y} + \dfrac{P_2}{l_x}\right) k_2^2 & 
        \dfrac{4k_o}{l_x l_y} k_1 k_2 \\[5mm]
        \dfrac{4k_o}{l_x l_y} k_1 k_2 & 
        \left(\dfrac{4k_o}{l_x l_y} + \dfrac{P_1}{l_y}\right) k_1^2 + \dfrac{k_y l_y}{l_x} k_2^2
        \end{bmatrix},
    \end{equation}
    revealing that {\it it represents the acoustic tensor of the equivalent material}, eq.~(\ref{acusticazzo2}), where now $\cos\gamma = k_1$ and $\sin \gamma=k_2$. 

    The continuum energy of the effective material is 
    \begin{equation}
        W^{\text{EFF}} = \frac{1}{4\pi^2} \int_{\fR^2} \frac{1}{2} 
        \begin{bmatrix} 
        \widehat{u}(\bk) & \widehat{v}(\bk)
        \end{bmatrix} 
        \cdot \bK_{0}^{\text{EFF}}(\bk) 
        \begin{bmatrix} 
        \overline{\widehat{u}(\bk)} \\[3mm] \overline{\widehat{v}(\bk)}
        \end{bmatrix} 
        \, d\bk, 
    \end{equation}
    which admits the representation
    \begin{equation}
        W^{\text{EFF}} = \frac{1}{4\pi^2} \int_{\fR^2} \frac{1}{2} 
        \begin{bmatrix} 
        i k_1 \widehat{u}(\bk) & i k_2 \widehat{v}(\bk)
        \end{bmatrix} 
        \cdot \fC_{0}^{\text{EFF}}
        \begin{bmatrix} 
        -i k_1 \overline{\widehat{u}(\bk)} \\[3mm] -i k_2 \overline{\widehat{v}(\bk)}
        \end{bmatrix} 
        \, d\bk, 
    \end{equation}
    providing the effective constitutive tensor $\fC_0^{\text{EFF}}$, which is found to coincide with the effective constitutive tensor $\mE$ given in eq.~\eqref{paola_gatti}.
    
    The conclusion is that the homogenization procedure introduced in Section \ref{modellazzo} is rigorous and that the technique based on variational calculus and Gamma-convergence is analogous to the dynamic asymptotic homogenization scheme developed in \cite{bordiga_dynamics_2021}. In fact, both these approaches lead to the acoustic tensor of the equivalent material.

\end{itemize}

\section{The proof-of-principle model: (PD) and (SE), shear bands, compaction bands, mixed modes, and restabilization}
\label{esempioni}

The conditions of positive definiteness of tensor $\mathcal{E}$, briefly (PD), eq.~\eqref{pd} and the equivalent eq.~\eqref{orchite}, is a local sufficient condition for uniqueness of the incremental response of the effective elastic material for a finite body subject to any mixed (imposed displacements and tractions) boundary conditions. For a  homogeneously deformed and stressed infinite body, the van Hove theorem \cite{vanhove, bigoni_nonlinear_2012} states that failure of (PD) does not imply any instability or bifurcation, but the relevant condition becomes the loss of (SE), occurring the first time when the determinant of the acoustic tensor eq.~(\ref{quartica}) vanishes from positive and thus representing also a failure of ellipticity (E). 
The latter case corresponds to the formation of localized deformation bands, which may be shear, compaction, or mixed bands. In the first two cases, the localized strain is either shear or normal to the band, respectively, while in the mixed case, the localized strain includes both shear and normal components.
Results presented in Section \ref{reticolazzo} show that the (PD) condition remains the same for the proof-of-principle model, while loss of (SE) in the solid corresponds to the emergence of a macrobifurcation in the lattice.  Note that the possibility of microbifurcations is excluded only when (PD) holds. 

In the following examples, regions of (PD) and (SE) pertaining to the equivalent continuum and to the infinite lattice have been found to coincide, so that in the lattice, the first bifurcation was always found to be macroscopic. 

On the introduction of the following dimensionless parameters 
\begin{equation}
 p_1 = \frac{P_1}{k_x l_x^0},
    \quad
    p_2 = \frac{P_2}{k_x l_x^0}, 
    \quad
    \xi = \frac{l_y^0}{l_x^0},
    \quad
    \kappa_x = \frac{k_x (l_x^0)^2}{k_o},
    \quad
    \kappa_y = \frac{k_y (l_x^0)^2}{k_o},
\end{equation}
both conditions (PD) and (SE) can be represented in the plane of biaxial stress $p_1$--$p_2$ as functions of the parameters $\xi$, $\kappa_x$, and $\kappa_y$,  defining the grid of axially deformable, but flexurally rigid bars. 

With the above dimensionless parameters, the requirement that the bars remain of finite length, in other words, that  $\lambda_1$ and $\lambda_2$ remain strictly positive, becomes
\begin{equation}
    p_1 > -1,
    \quad
    p_2 > - \frac{\kappa_y \xi}{\kappa_x},
\end{equation}
which limits the parameter space of the investigation. 

\begin{figure}[htbp]
	\centering
	{\phantomsubcaption\label{fig:stability_domains_cubic_1}}
	{\phantomsubcaption\label{fig:stability_domains_cubic_2}}
	{\phantomsubcaption\label{fig:stability_domains_cubic_5}}
	{\phantomsubcaption\label{fig:stability_domains_cubic_10}}
	{\phantomsubcaption\label{fig:stability_domains_cubic_20}}
	{\phantomsubcaption\label{fig:stability_domains_cubic_100}}
	\includegraphics[width=0.9\linewidth]{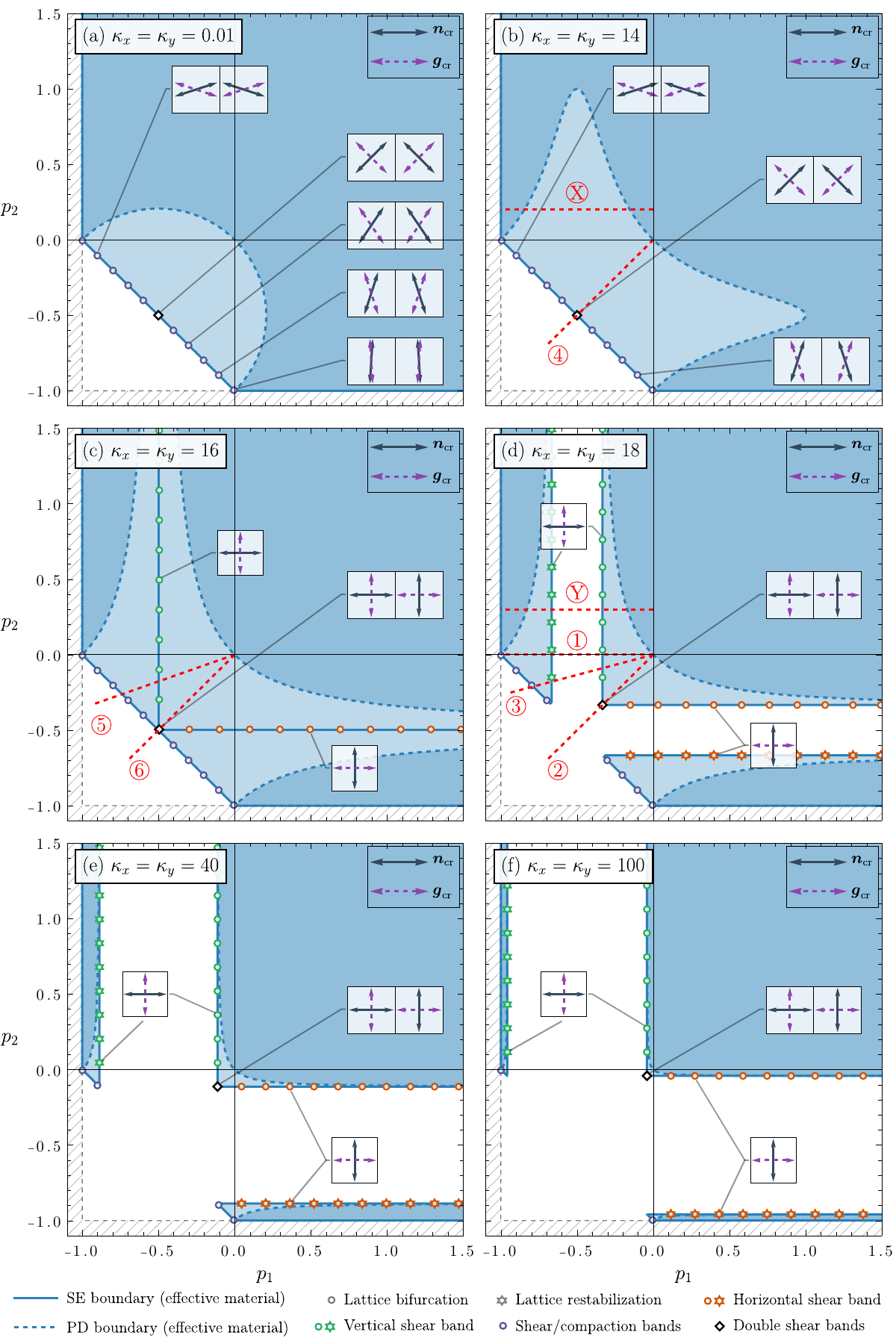}
	\caption{Strong ellipticity (SE), positive definiteness (PD), and lattice `first (macroscopic) bifurcation' domains for the \textit{cubic} grid with $\xi=1$, and six values of axial stiffness $\kappa_x=\kappa_y=\{0.01,7,8,9,20,100\}$. The arrows sketched in the insets represent the critical direction $\bn_{\text{cr}}$ and the associated mode $\bg_{\text{cr}}$ responsible for the loss of strong ellipticity. Failure of (PD) and subsequently of (SE) and their recovery are visible, for instance see path labelled \lq Y'. }
	\label{fig:stability_domains_cubic}
\end{figure}

\begin{figure}[htbp]
	\centering
	{\phantomsubcaption\label{fig:stability_domains_ortho_1}}
	{\phantomsubcaption\label{fig:stability_domains_ortho_2}}
	{\phantomsubcaption\label{fig:stability_domains_ortho_5}}
	{\phantomsubcaption\label{fig:stability_domains_ortho_10}}
	{\phantomsubcaption\label{fig:stability_domains_ortho_20}}
	{\phantomsubcaption\label{fig:stability_domains_ortho_100}}
	\includegraphics[width=0.9\linewidth]{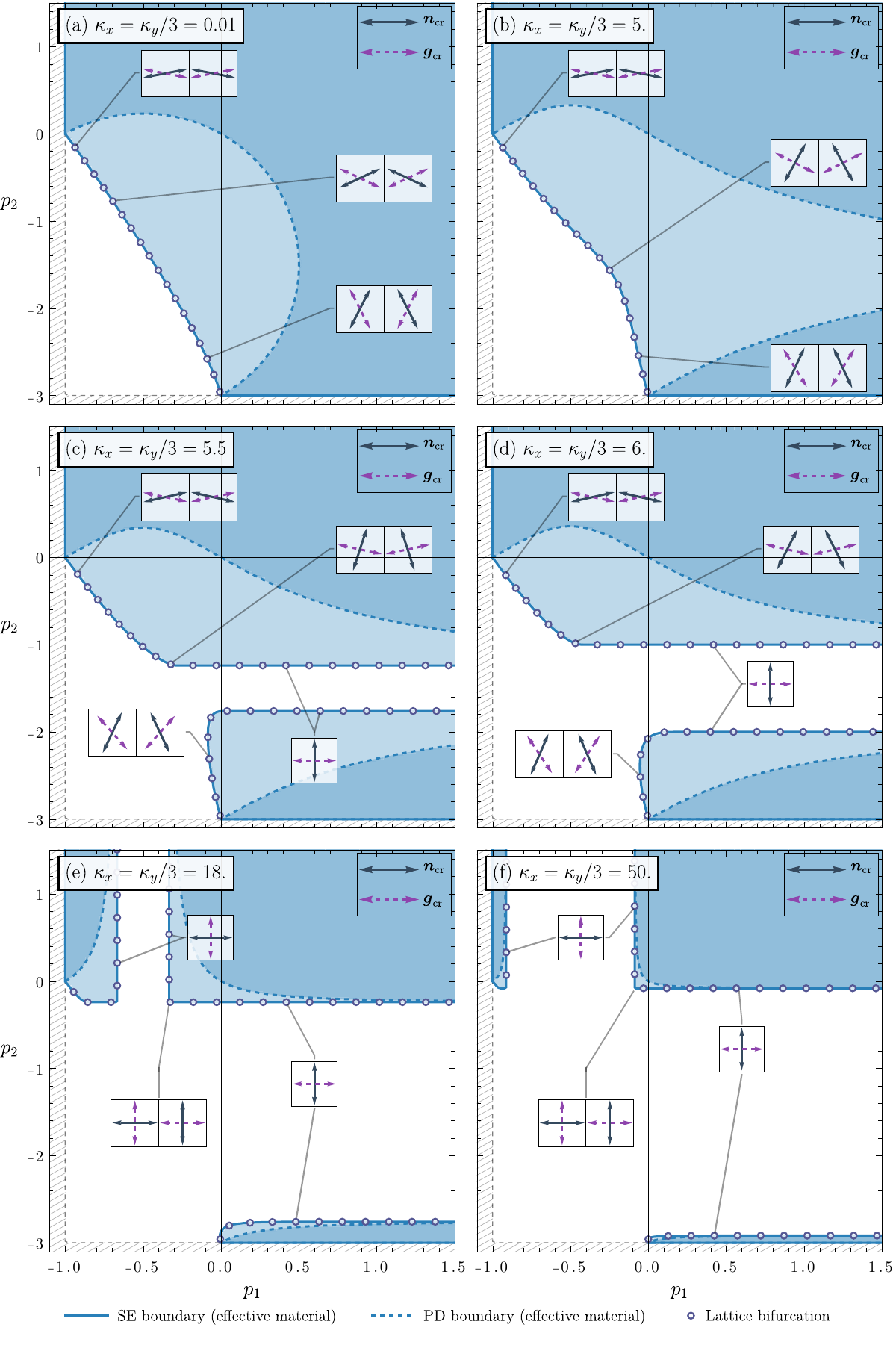}
	\caption{As for Fig.~\ref{fig:stability_domains_cubic}, except that now, the grid is orthotropic, because $\kappa_x=\kappa_y/3$, although $\xi=1$. Differently from the cubic case, the boundaries of (SE) are curved, and the localization bands are characterized by a mixed mode, ranging from shear bands to compaction bands. Restabilization remains evident.}
	\label{fig:stability_domains_ortho}
\end{figure}

A square grid, $\xi=1$ and $\kappa_x=\kappa_y$, and a rectangular grid, $\xi=1$ and $\kappa_x=\kappa_y/3$, of flexurally rigid elements are analyzed in Figs.~\ref{fig:stability_domains_cubic} and \ref{fig:stability_domains_ortho}, where the dark blue (light blue) domain represents PD (represents SE). Here, (SE) denotes strong ellipticity, occurring when the acoustic tensor is positive definiteness, so that failure of (SE) coincides with the first failure of (E). Increasing values of axial stiffnesses $\kappa_x$ and $\kappa_y$ are investigated. On the boundary of (SE), where the loss of ellipticity occurs, the insets report arrows representing the unit normal to the localization band, $\bn_{\text{cr}}$,  and the associated mode $\bg_{\text{cr}}$, so that the localization mode can be obtained from equation \eqref{modazzo}.

Fig.~\ref{fig:stability_domains_cubic}, referring to $\xi=1$ and $\kappa_x=\kappa_y=\{0.01,14,16,18,40,100\}$, shows the following features. 

\begin{itemize}

\item As expected, the (PD) region lies inside the (SE) domain. 

\item The material is cubic, so that the response is symmetric with respect to the diagonal of the map. 

\item For sufficiently small axial stiffnesses, $\kappa_x=\kappa_y= 0.01 \mbox{ and } 14$ in the figure, restabilization may only occur with regard to (PD) (path labelled \lq X'), but not in terms of (SE). The only difference between the two axial stiffnesses 0.01 and 14 is in the form and size of the (PD) domain, while (SE) does not change (which is also true for $\kappa_x=\kappa_y= 16$). The first bifurcation in the grid coincides with the loss of (SE), possible at a straight segment of its boundary segment aligned orthogonally to the symmetry axis. Only on the intersection point with the symmetry axis (see the paths labelled \lq 4' and \lq 6'), two symmetric shear bands are possible. Otherwise, the localization bands always contain a strain component orthogonal to the band and occur in a {\it mixed mode}. 

\item Restabilization in terms of (SE) corresponds to the formation of vertical and horizontal \lq channels' in the maps, clearly visible for $\kappa_x=\kappa_y = 18, 40, \text{ and } 100$. 

\item All boundaries of (SE) are piecewise linear; on vertical and horizontal segments, loss of (SE) occurs with shear band formation, while on the segment(s) inclined at 45$^\circ$ the localization bands are mixed mode, ranging from the extreme cases of shear (at intersections with the diagonal) and {\it compaction} bands (occurring in the limits $p_1=-1$ and $p_2=-1$. 

\item The mode of localization bands is independent of the stiffness $\kappa_x=\kappa_y$ of the grid, but only depends on the inclination of the (SE) boundary. Thus, the localization mode is the same along equally-inclined boundaries of (SE) for different values of $\kappa_x=\kappa_y$.  

\item The path labelled \lq 1' originates from the unloaded (stable) case, hits the boundary of (SE), then travels inside an unstable (white) region and hits the boundary of (SE) again at the instant of restabilization. At both boundaries, {\it two shear bands} occur, with $\bn_{\text{cr}}\cdot \bg_{\text{cr}} = 0$.

\item The path labelled \lq 3' shows a loss of (SE) and a recovery, corresponding in both cases to a shear band, and a final failure of (SE) with the formation of mixed mode bands. Restabilization does not occur for the path labelled \lq 2'. 

\item The path labelled \lq Y' starts from a (PD) state $\{p_1=0, p_2>0 \}$, so that a loss of (PD) is observed before the loss of (SE) occurring with shear band formation, while (SE) is subsequently recovered and later (PD).

\item The special case $\kappa_x=\kappa_y = 16$ corresponds to the nucleation of the instability channels, which are only visible as two straight lines. In this very particular case, (SE) is lost and immediately recovered (path labelled \lq 5'), a strange situation, for which it is difficult to provide a mechanical interpretation. 

\item A general conclusion is that an increase in axial stiffness is a destabilizing factor, promoting the loss of (PD) and (SE), and delaying their recovery.
In other words, {\it stiffening may promote material instability}, a statement that may seem paradoxical, but is consistent with analogous results found in the structures analysed in \cite{tarnai_destabilizing_1980, feodosyev_selected_1977, potier-ferry_foundations_1987, koutsogiannakis_double_2023}. 
    
\end{itemize}

Data referred to the paths reported in red in Fig.~\ref{fig:stability_domains_cubic} (inclined at the angle $\phi$ with respect to the positive $p_1$ axis) are provided in Table~\ref{tab:cases_analyzed}. The table supplies the values of biaxial loads $\bp_E$ for loss of (SE) and the corresponding inclination angles for the localization band, $\theta_\bn$, and mode $\theta_\bg$ . 


\begin{table}[htb!]
    \centering
    \begin{tabular}{@{}llllll@{}}
        \toprule
        $\kappa$    & Path        & $\phi$        & (SE) loss                         & (SE) recovery             & (SE) re-loss 
        \\ \midrule               
        $18$        & \Circled{1} & $0^\circ$     & $\bp_E=\{-0.333,0\}$             & $\bp=\{-0.667,0\}$       & -              \\
                    &             &               & $\theta_\bn=0^\circ$             & $\theta_\bn=0^\circ$     & -              \\
                    &             &               & $\theta_\bg=90^\circ$            & $\theta_\bg=90^\circ$    & -              \\
                    \cmidrule(l){2-6}
                    & \Circled{2} & $45^\circ$    & $\bp_E=\{-0.333,-0.333\}$        & -                        & -              \\
                    &             &               & $\theta_\bn=0^\circ,90^\circ$    & -                        & -              \\
                    &             &               & $\theta_\bg=90^\circ,0^\circ$    & -                        & -              \\
                    \cmidrule(l){2-6}
                    & \Circled{3} & $15^\circ$    & $\bp_E=\{-0.333,-0.089\}$        & $\bp=\{-0.667,-0.179\}$  & $\bp=\{-0.789,-0.211\}$             \\
                    &             &               & $\theta_\bn=0^\circ$             & $\theta_\bn=0^\circ$     & $\theta_\bn=27.4^\circ,152.6^\circ$ \\
                    &             &               & $\theta_\bg=90^\circ$            & $\theta_\bg=90^\circ$    & $\theta_\bg=152.6^\circ,27.4^\circ$ \\ 
                    \midrule               
        $14$        & \Circled{4} & $45^\circ$    & $\bp_E=\{-0.5,-0.5\}$            & -                        & -              \\
                    &             &               & $\theta_\bn=45^\circ,135^\circ$  & -                        & -              \\
                    &             &               & $\theta_\bg=135^\circ,45^\circ$  & -                        & -              \\ 
                    \midrule               
        $16$        & \Circled{5} & $19.5^\circ$  & $\bp_E=\{-0.5,-0.177\}$          & $\bp=\{-0.5,-0.177\}$    & $\bp=\{-0.738,-0.262\}$             \\
                    &             &               & $\theta_\bn=0^\circ$             & $\theta_\bn=0^\circ$     & $\theta_\bn=30.8^\circ,149.2^\circ$ \\
                    &             &               & $\theta_\bg=90^\circ$            & $\theta_\bg=90^\circ$    & $\theta_\bg=149.2^\circ,30.8^\circ$ \\     
                    \cmidrule(l){2-6}
                    & \Circled{6} & $45^\circ$    & $\bp_E=\{-0.5,-0.5\}$            & -                        & -              \\
                    &             &               & $\theta_\bn=\forall\theta\in[0^\circ,\pi^\circ]$  & -       & -              \\
                    &             &               & $\theta_\bg\perp\theta_\bn$      & -                        & -              \\
        \bottomrule
    \end{tabular}
    \caption{Loss, recovery, and re-loss of strong ellipticity for different axial stiffnesses $\kappa_x=\kappa_y=\kappa$ and the radial loading paths $\bp=\{p_1,p_2\}=p\{\cos\phi,\sin\phi\}$ reported red in Fig.~\ref{fig:stability_domains_cubic}. The radial loading path in the $p_1$-$p_2$ plane is singled out by the angle $\phi$ measured counterclockwise from the $p_1$ axis.}
    \label{tab:cases_analyzed}
\end{table}

Fig.~\ref{fig:stability_domains_ortho}, referring to $\xi=1$ and $\kappa_x=\kappa_y/3=\{0.1,5,5.5,6,18,50\}$, shows a situation similar to that reported in Fig.~\ref{fig:stability_domains_cubic}, except that now the material is orthotropic and the diagonal symmetry is lost. 
Orthotropy is not geometrical, because $\xi=1$, but is related to the difference between the stiffnesses of vertical and longitudinal springs. 
The following features are remarkable and cannot be observed in the previously analyzed case of cubic symmetry.  

\begin{itemize}

\item The boundaries of (SE) are all curved, but may present straight portions.

\item At increasing axial stiffness, the boundaries of (PD) and (SE) tend to coincide. The latter tend to become composed of straight segments. 

\item Along straight boundaries of (SE), loss of (SE) occurs in the form of shear bands, while mixed modes prevail on curved boundaries. 

\item Note the twin shear bands indicated in the figure corresponding to  $\kappa_x=\kappa_y/3=5$, which have opposite inclinations, different from the vertical and horizontal directions defining the rods forming the grid.
    
\end{itemize}

\begin{figure}[htb!]
    \centering
    \begin{subfigure}{0.30\textwidth}
        \centering
        \caption{\label{fig:bsurf1}Path \Circled{1}}
        \includegraphics[width=0.90\linewidth]{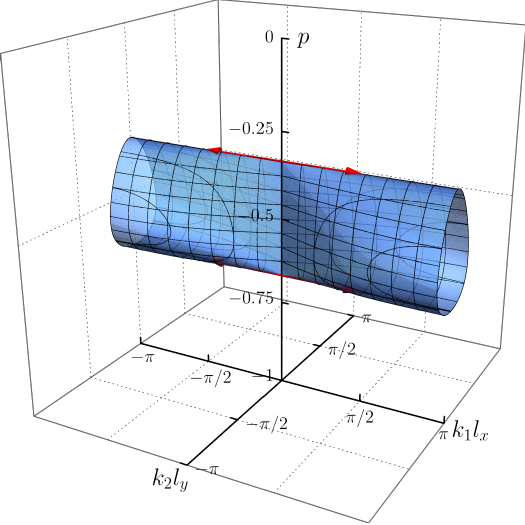}
    \end{subfigure}
    \begin{subfigure}{0.30\textwidth}
        \centering
        \caption{\label{fig:bsurf2}Path \Circled{2}}
        \includegraphics[width=0.90\linewidth]{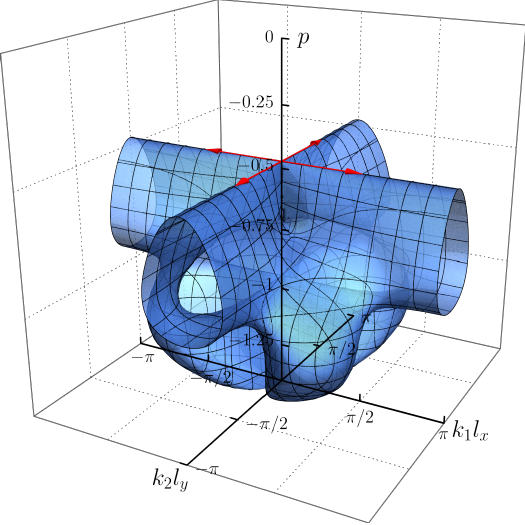}
    \end{subfigure}
    \begin{subfigure}{0.30\textwidth}
        \centering
        \caption{\label{fig:bsurf3}Path \Circled{3}}
        \includegraphics[width=0.90\linewidth]{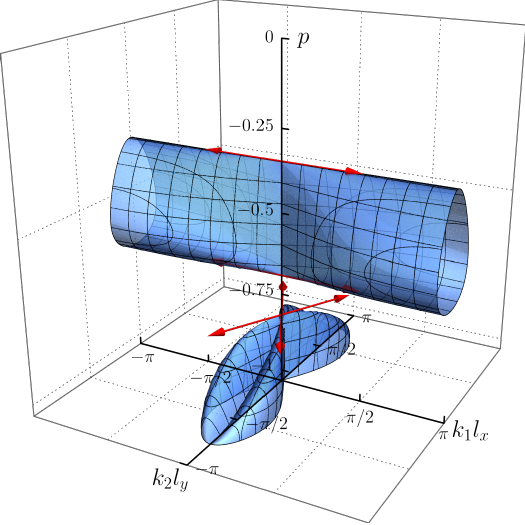}
    \end{subfigure} \\ 
    \vspace{4mm}
    \begin{subfigure}{0.30\textwidth}
        \centering
        \caption{\label{fig:bsurf4}Path \Circled{4}}
        \includegraphics[width=0.90\linewidth]{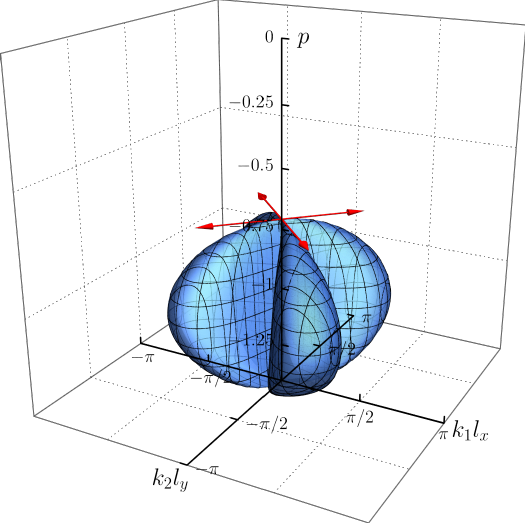}
    \end{subfigure}
    \begin{subfigure}{0.30\textwidth}
        \centering
        \caption{\label{fig:bsurf5}Path \Circled{5}}
        \includegraphics[width=0.90\linewidth]{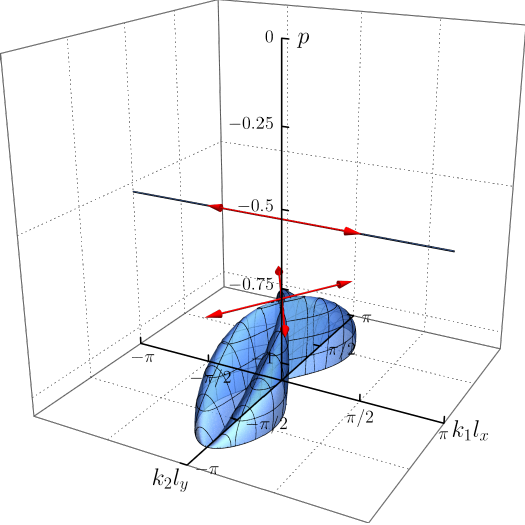}
    \end{subfigure}
    \begin{subfigure}{0.30\textwidth}
        \centering
        \caption{\label{fig:bsurf6}Path \Circled{6}}
        \includegraphics[width=0.90\linewidth]{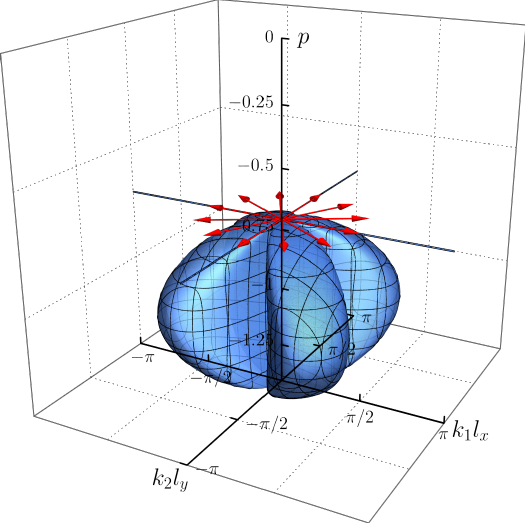}
    \end{subfigure}
    \caption{Bifurcation surfaces obtained by plotting the condition \eqref{eq:det} in the space $\{k_1 l_x, k_2 l_y, p\}$, referred to the specific radial loading paths $p\{\cos\phi,\sin\phi\}$ shown as red dashed lines in Fig.~\ref{fig:stability_domains_cubic}. The red arrows denote the bifurcation macro-modes, which are observed when $k_1=k_2=0$. Restabilization occurs in paths labelled \lq 1', \lq 3', and \lq 5'. Note that the first bifurcation along paths 1-3 and the restabilization along path 3 correspond to the simultaneous occurrence of both a macro mode and infinite microscopic modes, as the mode wavelength is arbitrary.}
    \label{fig:bsurf}
\end{figure}

In the closure of this section, bifurcation surfaces are reported in Fig.~\ref{fig:bsurf}, referred to the space $\{\tilde{k}_1, \tilde{k}_2, p\}$, where $\tilde{k}_1 = k_1 l_x$ and $\tilde{k}_2 = k_2 l_y$ are normalised wavenumbers. The surfaces are obtained for the grid of bars satisfying cubic symmetry. The bifurcation analysis was performed via the Floquet-Bloch representation of the lattice displacement through equation \eqref{eq:det} and refer to the radial loading paths $p\{\cos\phi,\sin\phi\}$ shown as red dashed lines in Fig.~\ref{fig:stability_domains_cubic}. 
 
The surfaces correspond to the satisfaction of the bifurcation condition, eq.~\eqref{eq:det}, so that for all other points, the incremental solution is unique, although may be unstable. The radial paths start from the unloaded state, $p=0$, from which compression is increased, and the loading parameter $p<0$ decreases. 
The red arrows in the graphs denote the direction of eigenvectors, which correspond to the direction of the normal to the localization band. The macroscopic modes of bifurcation occur for vanishing wavevector, $\tilde{k}_1=\tilde{k}_2=0$, while microscopic modes are characterized by values of $\tilde{k}_i \neq 0$. 

\begin{itemize}

\item The path labelled \lq 1' shows a first bifurcation, occurring simultaneously in a macroscopic mode and in infinite microscopic modes (because the surface has a horizontal line) with $\tilde{k}_1\neq 0$ and $\tilde{k}_2=0$. The macroscopic mode corresponds to a shear band formation in the equivalent material with the normal aligned parallel to the $\tilde{k}_1$-axis. Restabilization is finally observed along the path (because the surface is closed), following another point where multiple bifurcations and shear bands occur. 

\item Also for the path labelled \lq 2', the first bifurcation simultaneously involves a macroscopic mode together with infinite microscopic modes (because the surface has two orthogonal horizontal lines), with either  $\tilde{k}_1\neq 0$ and $\tilde{k}_2=0$ or $\tilde{k}_1= 0$ and $\tilde{k}_2 \neq 0$. Restabilization does not occur in this case, in the sense that it would become possible when the length of the bars vanishes, a situation which is not admissible. 

\item The path labelled \lq 3' is similar to path \lq 1', but after the initial bifurcation and restabilization, a final bifurcation occurs, corresponding to mixed mode localization bands.

\item Restabilization is excluded for paths labelled, \lq 2', \lq 4', and \lq 6', because the surface is punctured only one time, but in the two cases the bifurcation modes are different. Infinite shear bands become possible when loss of (SE) occurs for path labelled \lq 6', due to the high symmetry. 

\item The path labelled \lq 5' corresponds to the strange behaviour where strong ellipticity is lost at a point, immediately recovered, and finally lost again. 

\end{itemize}

All features already pointed out for Fig.~\ref{fig:stability_domains_cubic} are confirmed from the analysis of the bifurcation surfaces. 

\clearpage

\section{Islands of instability: a grid of axially and flexurally deformable elastic rods}
\label{isolotti}

\begin{figure}[htbp]
    \centering
    \begin{subfigure}{0.45\textwidth}
        \centering        
        \includegraphics[height=0.65\linewidth]{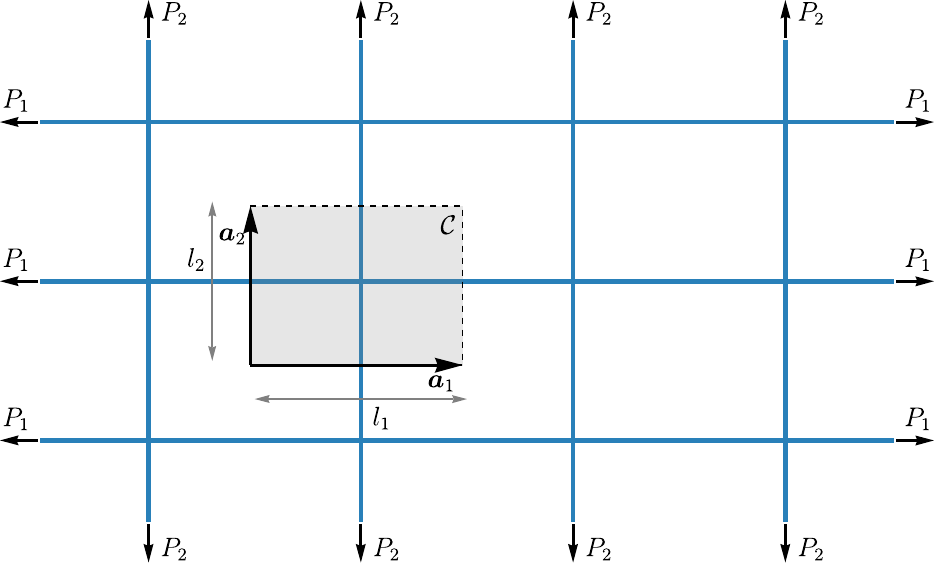}
    \end{subfigure}%
    \hspace{3mm}
    \begin{subfigure}{0.45\textwidth}
        \centering
        \includegraphics[height=0.65\linewidth]{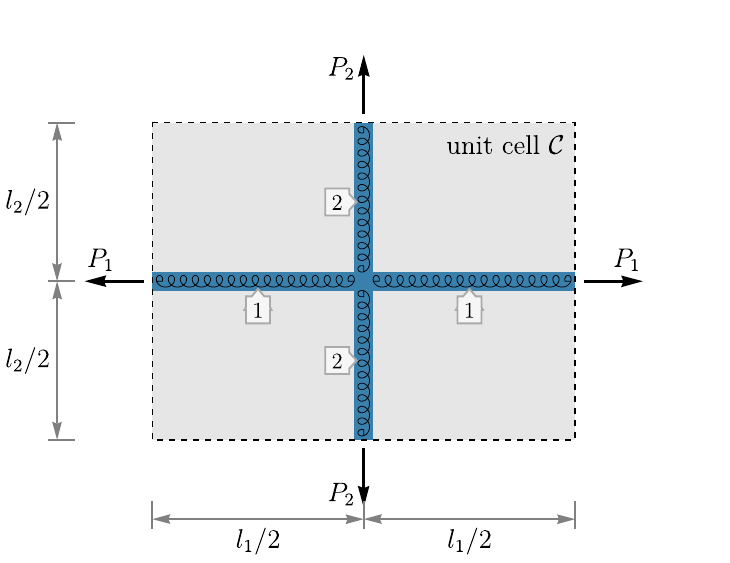}
    \end{subfigure}%
    \caption{Left: a rectangular lattice of axially and flexurally deformable rods realizes an architected material capable of losing and regaining ellipticity under monotonously increasing compressive load. Right: the unit cell; axially, the rods behave as linear springs, thus providing a highly compliant axial deformation within a linear range.}
    \label{fig:sliding_grid}
\end{figure}

The elastic structure with lumped degrees of freedom has shown its merits, highlighting through homogenization the phenomenon of restabilization and providing a deep insight into material instability. Now, a new model based on axially and flexurally deformable elastic rods, defining the periodic grid shown in Fig.~\ref{fig:sliding_grid}, is used to confirm findings based on the lumped structure and disclose further features of interest. 
In particular, it is shown below that the \lq channels' may become \lq islands' of instability in a \lq sea' of stability. Moreover, it is also proven that structures can be designed to exhibit radial paths of loading, emanating from the unloaded state, that may graze instability regions. In the latter circumstance, it is shown that a perturbation approach evidences the formation of different forms of shear bands at different loadings, occurring subsequently while the loading path remains inside the (SE) region. 

The elastic rods are equipped with the same bending stiffness in the horizontal and vertical directions of the grid
\begin{equation}
    B_x = B_y = B, 
\end{equation} 
while the axial stiffnesses of the horizontal and vertical rods, $A_x$ and $A_y$, are assumed equivalent to those of linear elastic springs, to allow an easier comparison with the proof-of-concept model with lumped degrees of freedom
\begin{equation}
    A_x = k_x l_x, \quad A_y = k_y l_y.
\end{equation}
%

The constitutive tensor of the equivalent elastic material is obtained with the procedure described in \cite{bordiga_dynamics_2021}. The components of the  elasticity tensor, prestressed as a function of the loads $P_1$ and $P_2$ acting on the grid, are
\begin{equation}
\label{eq:ctens1}
\begin{aligned}
    \displaystyle
    \mathcal{E}_{1111} &= \frac{k_x l_x}{l_y}, \qquad \mathcal{E}_{2222} = \frac{k_y l_y}{l_x}
    \qquad 
    \mathcal{E}_{1221} = \mathcal{E}_{2112} = \frac{2}{\Delta} E_1^- E_2^- \sqrt{B} P_1 P_2, \\[4mm]
    \mathcal{E}_{1212} &= \frac{l_y}{l_x \Delta} \Big[E_1^- E_2^+ l_x P_1 P_2^{3/2} - 2 E_1^- E_2^- \sqrt{B} P_2^2 + E_1^+ E_2^- l_x P_1^{1/2} P_2^{2} \Big], \\[4mm]
    \mathcal{E}_{2121} &= \frac{l_x}{l_y \Delta} \Big[E_1^+ E_2^- l_y P_1^{3/2} P_2 - 2 E_1^- E_2^- \sqrt{B} P_1^2 + E_1^- E_2^+ l_y P_1^{2} P_2^{1/2} \Big], 
\end{aligned}
\end{equation}
where
\begin{equation}
\begin{aligned}
    E_1^\pm &= \exp \left(l_x \sqrt{\frac{P_1}{B}}\right) \pm 1, \qquad E_2^\pm = \exp \left(l_y \sqrt{\frac{P_2}{B}}\right) \pm 1, \\[3mm]
    \Delta &= E_1^- E_2^+ l_x l_y P_1 P_2^{1/2} - 2 E_1^- E_2^- \sqrt{B} (P_1 l_x + P_2 l_y) + E_1^+ E_2^- l_x l_y P_1^{1/2} P_2.
\end{aligned}
\end{equation}

The elasticity tensor, eq.~\eqref{eq:ctens1}, can be rewritten in terms of the principal components of Cauchy stress, $T_1$ and $T_2$, as
\begin{equation}
\label{eq:ctens2}
\begin{aligned}
    \displaystyle
    \mathcal{E}_{1111} &= \frac{k_x l_x}{l_y}, \qquad \mathcal{E}_{2222} = \frac{k_y l_y}{l_x}, 
    \qquad
    \mathcal{E}_{1221} = \mathcal{E}_{2112} = \frac{2}{\Delta} E_1^- E_2^- \sqrt{B} T_1 T_2,
    \\[4mm]
    \mathcal{E}_{1212} &= \frac{1}{\Delta} \Big[E_1^- E_2^+ l_x^{1/2} l_y T_1 T_2^{3/2} - 2 E_1^- E_2^- \sqrt{B} T_2^2 + E_1^+ E_2^- l_x l_y^{1/2} T_1^{1/2} T_2^{2} \Big], \\[4mm]
    \mathcal{E}_{2121} &= \frac{1}{\Delta} \Big[E_1^+ E_2^- l_x l_y^{1/2} T_1^{3/2} P_2 - 2 E_1^- E_2^- \sqrt{B} T_1^2 + E_1^- E_2^+ l_x^{1/2} l_y T_1^{2} T_2^{1/2} \Big], 
\end{aligned}
\end{equation}
where
\begin{equation}
\begin{aligned}
    E_1^\pm &= \exp \left(l_x \sqrt{\frac{T_1 l_y}{B}}\right) \pm 1, \qquad E_2^\pm = \exp \left(l_y \sqrt{\frac{T_2 l_x}{B}}\right) \pm 1, \\[4mm]
    \Delta &= E_1^- E_2^+ l_x^{1/2} l_y T_1 T_2^{1/2} - 2 E_1^- E_2^- \sqrt{B} (T_1 + T_2) + E_1^+ E_2^- l_x l_y^{1/2} T_1^{1/2} T_2.
\end{aligned}
\end{equation}

\begin{figure}[htbp]
	\centering
	{\phantomsubcaption\label{fig:stability_domains_cubic_deform_1}}
	{\phantomsubcaption\label{fig:stability_domains_cubic_deform_2}}
	{\phantomsubcaption\label{fig:stability_domains_cubic_deform_5}}
	{\phantomsubcaption\label{fig:stability_domains_cubic_deform_10}}
	{\phantomsubcaption\label{fig:stability_domains_cubic_deform_20}}
	{\phantomsubcaption\label{fig:stability_domains_cubic_deform_100}}
	\includegraphics[width=0.9\linewidth]{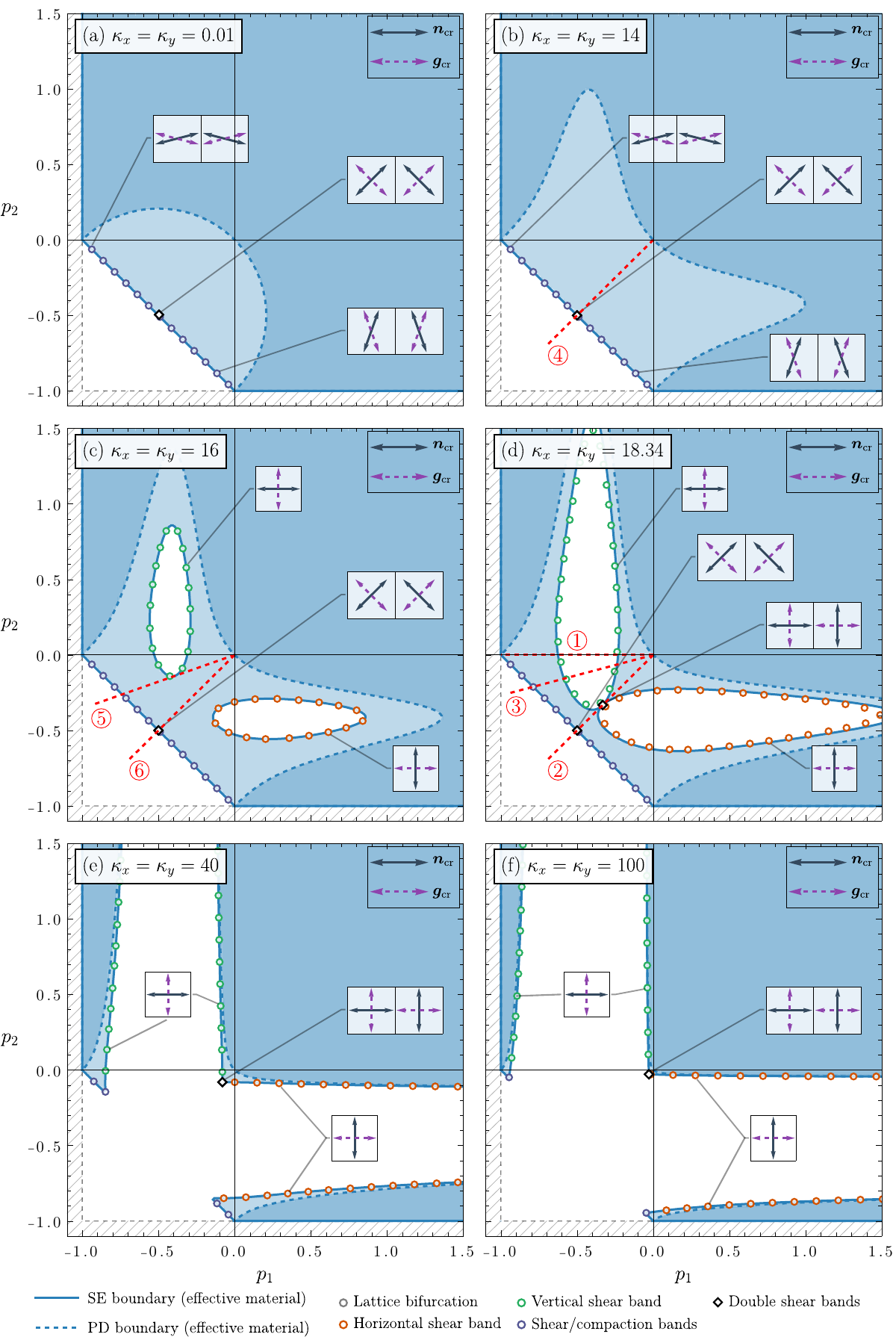}
	\caption{Stability domains as for Fig.~\ref{fig:stability_domains_cubic}, except that now the grid comprises flexurally and axially deformable elastic rods. Note the \lq islands of instability in the sea of stability (SE)'.}
\label{fig:stability_domains_cubic_deform}
\end{figure}

The following dimensionless parameters are introduced 
\begin{equation}
    \xi = \frac{l_y^0}{l_x^0},
    \quad
    \kappa_x = \frac{k_x (l_x^0)^3}{2 B},
    \quad
    \kappa_y = \frac{k_y (l_x^0)^3}{2 B},
    \quad
    p_1 = \frac{P_1}{k_x l_x^0},
    \quad
    p_2 = \frac{P_2}{k_x l_x^0}, 
\end{equation}
and results in terms of (PD) and (SE) domains in the $p_1$--$p_2$ plane are reported in Fig.~\ref{fig:stability_domains_cubic_deform}. 

The figure has to be compared with the analogue for the lumped degrees of freedom structure, Fig.~\ref{fig:stability_domains_cubic}. The numerical values of the parameters are the same (except for $\kappa=18.34$) in the two figures, namely, $\kappa_x=\kappa_y=\{0.01,14,16,18.34,40,100\}$ and $\xi=1$. Moreover, the radial paths (shown in red) are also the same, but now lead to the values reported in Table~\ref{tab:cases_analyzed_d}. 

The stability maps plotted in the figure agree well with the proof-of-principle structure and confirm all previous findings. However, the two following features are important and now disclosed.
\begin{itemize}
\item Cases $\kappa_x=\kappa_y = 16$ and $18.34$ show the existence of \lq islands' of instability, where (SE) is violated, in a sea of \lq stability', where (SE) holds.
\item The radial paths labelled \lq 2' and \lq 5' are tangent respectively to two or one island of loss of (SE). In this situation, a perturbation applied when the material is sufficiently close to instability reveals 
the appearance, disappearance, and reappearance of strain localization. The latter appears two times, the first \lq near the island' and the second near the \lq final frontier', a feature sketched in Figs.~\ref{figata} and \ref{figata_2}. The discussion on this important point is deferred to Section \ref{ritornano} below. 
\end{itemize}


\begin{table}[htb!]
    \centering
    \begin{tabular}{@{}llllll@{}}
        \toprule
        $\kappa$    & Path        & $\phi$       & (E) loss                         & (E) recovery               & (E) re-loss 
        \\ \midrule               
        $18.34$     & \Circled{1} & $0^\circ$    & $\bp_E=\{-0.231,0\}$             & $\bp=\{-0.625,0\}$         & -              \\
                    &             &              & $\theta_\bn=0^\circ$             & $\theta_\bn=0^\circ$       & -              \\
                    &             &              & $\theta_\bg=90^\circ$            & $\theta_\bg=90^\circ$      & -              \\
                    \cmidrule(l){2-6}
                    & \Circled{2} & $45^\circ$   & $\bp_E=\{-0.333,-0.333\}$        & $\bp=\{-0.333,-0.333\}$    & $\bp=\{-0.5,-0.5\}$                \\
                    &             &              & $\theta_\bn=0^\circ$             & $\theta_\bn=0^\circ$       & $\theta_\bn=45^\circ,135^\circ$    \\
                    &             &              & $\theta_\bg=90^\circ$            & $\theta_\bg=90^\circ$      & $\theta_\bg=135^\circ,45^\circ$    \\
                    \cmidrule(l){2-6}
                    & \Circled{3} & $15^\circ$   & $\bp_E=\{-0.238,-0.064\}$        & $\bp=\{-0.586,-0.157\}$    & $\bp=\{-0.789,-0.211\}$             \\
                    &             &              & $\theta_\bn=0^\circ$             & $\theta_\bn=0^\circ$       & $\theta_\bn=27.4^\circ,152.6^\circ$ \\
                    &             &              & $\theta_\bg=90^\circ$            & $\theta_\bg=90^\circ$      & $\theta_\bg=152.6^\circ,27.4^\circ$ \\ 
                    \midrule               
        $14$        & \Circled{4} & $45^\circ$   & $\bp_E=\{-0.5,-0.5\}$            & -                          & -              \\
                    &             &              & $\theta_\bn=45^\circ,135^\circ$  & -                          & -              \\
                    &             &              & $\theta_\bg=135^\circ,45^\circ$  & -                          & -              \\ 
                    \midrule               
        $16$        & \Circled{5} & $19.5^\circ$ & $\bp_E=\{-0.393,-0.139\}$        & $\bp=\{-0.393,-0.139\}$    & $\bp=\{-0.739,-0.261\}$             \\
                    &             &              & $\theta_\bn=0^\circ$             & $\theta_\bn=0^\circ$       & $\theta_\bn=30.8^\circ,149.2^\circ$ \\
                    &             &              & $\theta_\bg=90^\circ$            & $\theta_\bg=90^\circ$      & $\theta_\bg=149.2^\circ,30.8^\circ$ \\     
                    \cmidrule(l){2-6}
                    & \Circled{6} & $45^\circ$   & $\bp_E=\{-0.5,-0.5\}$            & -                          & -              \\
                    &             &              & $\theta_\bn=45^\circ,135^\circ$  & -                          & -              \\
                    &             &              & $\theta_\bg=135^\circ,45^\circ$  & -                          & -              \\
        \bottomrule
    \end{tabular}
    \caption{As for Table~\ref{tab:cases_analyzed}, except that now the grid comprises flexurally and axially deformable elastic rods. The radial paths numbered 1--6 are referred to Fig.~\ref{fig:stability_domains_cubic_deform}, where they are drawn as red dashed lines.}
    \label{tab:cases_analyzed_d}
\end{table}

\begin{figure}[htb!]
    \centering
    \begin{subfigure}{0.30\textwidth}
        \centering
        \caption{\label{fig:bsurfd1}Path \Circled{1}}
        \includegraphics[width=0.90\linewidth]{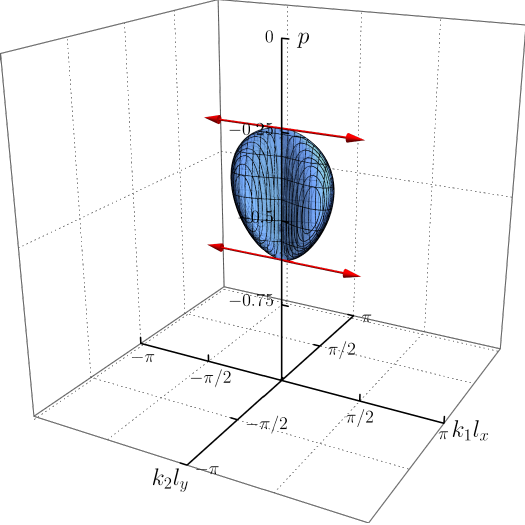}
    \end{subfigure}
    \begin{subfigure}{0.30\textwidth}
        \centering
        \caption{\label{fig:bsurfd2}Path \Circled{2}}
        \includegraphics[width=0.90\linewidth]{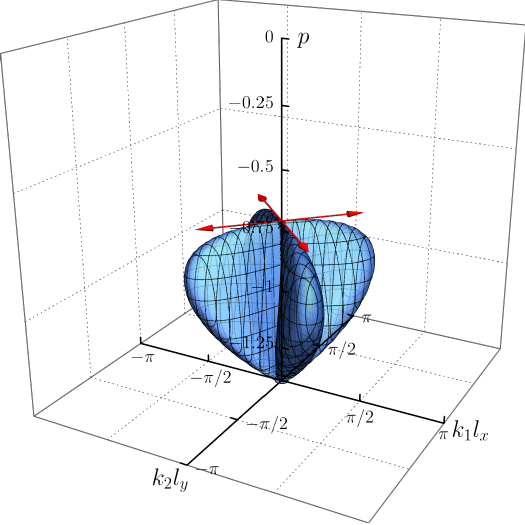}
    \end{subfigure}
    \begin{subfigure}{0.30\textwidth}
        \centering
        \caption{\label{fig:bsurfd3}Path \Circled{3}}
        \includegraphics[width=0.90\linewidth]{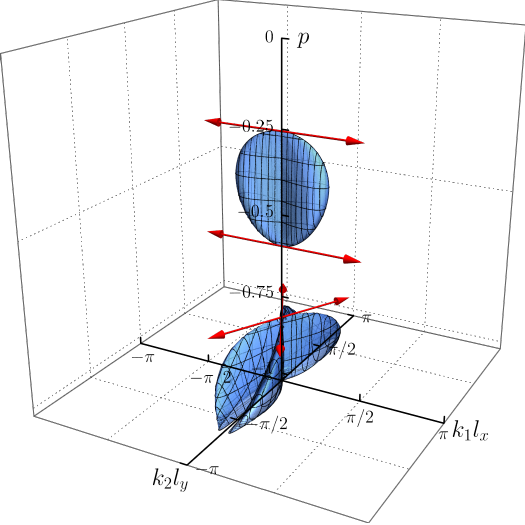}
    \end{subfigure} \\ 
    \vspace{4mm}
    \begin{subfigure}{0.30\textwidth}
        \centering
        \caption{\label{fig:bsurfd4}Path \Circled{4}}
        \includegraphics[width=0.90\linewidth]{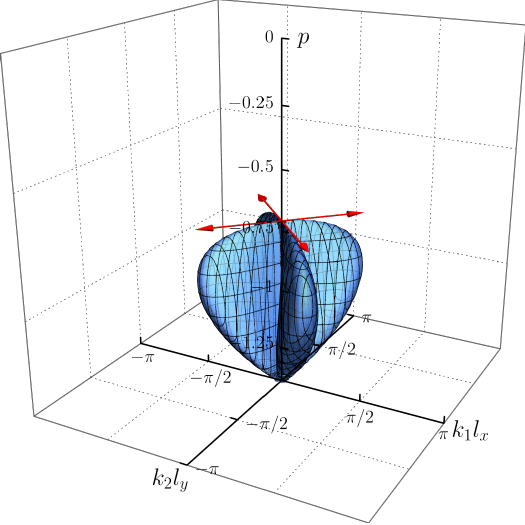}
    \end{subfigure}
    \begin{subfigure}{0.30\textwidth}
        \centering
        \caption{\label{fig:bsurfd5}Path \Circled{5}}
        \includegraphics[width=0.90\linewidth]{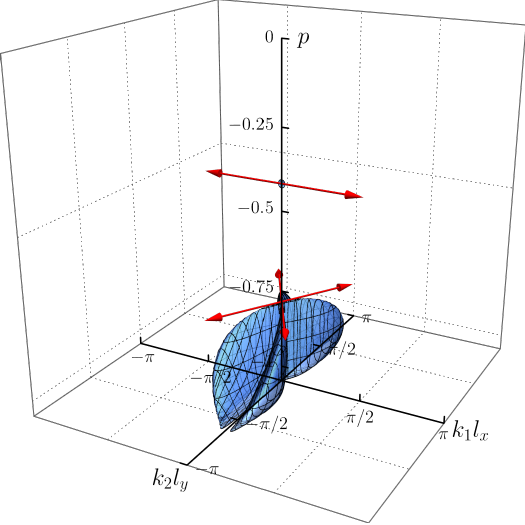}
    \end{subfigure}
    \begin{subfigure}{0.30\textwidth}
        \centering
        \caption{\label{fig:bsurfd6}Path \Circled{6}}
        \includegraphics[width=0.90\linewidth]{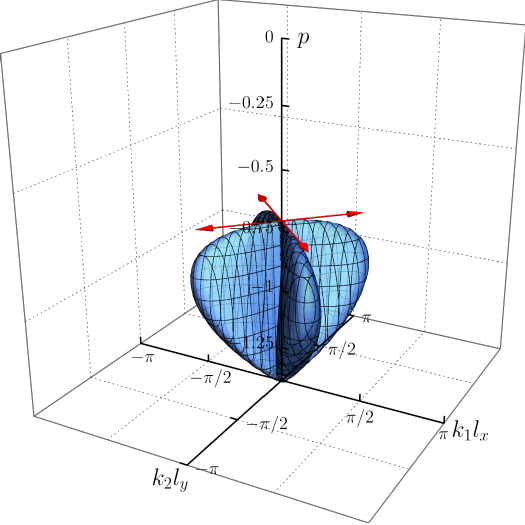}
    \end{subfigure}
    \caption{As for Fig.~\ref{fig:bsurf}, except that now the grid is made up of flexurally and axially deformable elastic rods. The bifurcation surfaces are referred to the specific radial loading paths shown as red dashed lines in Fig.~\ref{fig:stability_domains_cubic}. Note that the first bifurcation, the restabilization, and the final bifurcation involve only macroscopic modes.}
    \label{fig:bsurfd}
\end{figure}

Fig.~\ref{fig:bsurfd} reports the bifurcation surfaces similarly to the discrete case, Fig.~\ref{fig:bsurf}, so that the results presented in the two figures can be compared. The following features emerge from the comparison.
\begin{itemize}
\item The first bifurcation, the restabilization, and (when it occurs) the final bifurcation always occur in the macroscopic mode, so that the equivalent material loses, regains, and loses again ellipticity.
\item Mixed mode strain localization occurs only at restabilization; in all the other cases, shear bands occur at the loss of (SE). 
\item The path labelled \lq 6' does not show the occurrence of infinite shear bands, but only two, one vertical and one horizontal, are formed. 
\item In the path labelled \lq 5', the instability island is only touched, where a shear band occurs, together with the macroscopic bifurcation of the lattice.
\end{itemize}

\begin{figure}[htbp]
	\centering
	\includegraphics[width=0.80\linewidth]{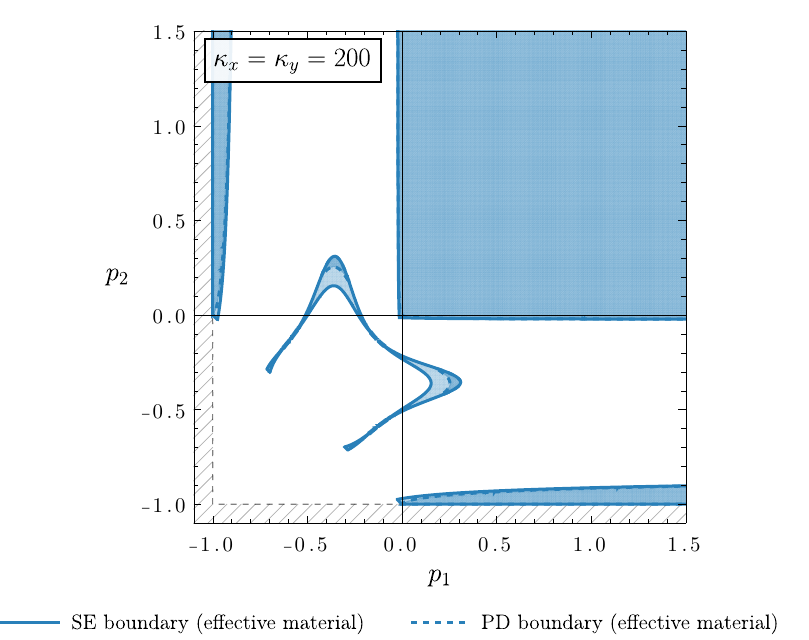}
	\caption{An island of restabilization in terms of (PD) and (SE) emerges at the intersection of two channels of instability. The restabilization is a genuine restabilization for the equivalent solid, which falsely represents the elastic grid, remaining unstable and characterized by an indefinite stiffness matrix.}
    \label{fig:stability_domains_cubic_deform_200}
\end{figure}

In the closure of this section, it is noted that, for sufficiently high values of $\kappa_x=\kappa_y$, islands of restabilization of the equivalent material may appear without the lattice being stable. This is a true restabilization for the solid, which falsely represents the behaviour of the elastic grid, remaining unstable. This behaviour, discovered already in \cite{bordiga_tensile_2022}, is shown in Fig.~\ref{fig:stability_domains_cubic_deform_200} for $\kappa_x=\kappa_y = 200$. 

Here, the situation is similar to that reported in Fig.~\ref{fig:stability_domains_cubic_deform}, except that now, inside the two \lq channels' of instability, an island of stability for the equivalent solid emerges. It can be observed that the restabilization occurs both in terms of (PD), appearing on the \lq shores' of the island, and of (SE), occupying the inside. 

The island visible in Fig.~\ref{fig:stability_domains_cubic_deform_200} provides a counterexample demonstrating that the (PD) condition in the solid may {\it not} coincide with the positive definiteness of the stiffness matrix of the elastic grid, appropriately reduced to eliminate rigid-body translations. In fact, the stiffness matrix for the grid is indefinite within the island where (PD) holds for the solid, confirming that the underlying lattice is unstable. 
The reason why the coincidence is verified for the proof-of-principle model and not for the elastic grid lies in the fact that the former is significantly less deformable than the latter. Specifically, the elastic grid, composed of flexurally deformable rods, admits periodic bifurcation modes that do not produce macroscopic strain, i.e. these modes correspond to null mean strain and stress. Such modes are impossible in the structure with lumped degrees of freedom, a feature which may inspire new applications for architected materials.


\subsection{The appearance, disappearance, and reappearance of shear bands}
\label{ritornano}

\begin{figure}[hbt!]
    \centering
    \includegraphics[width=0.98\linewidth]{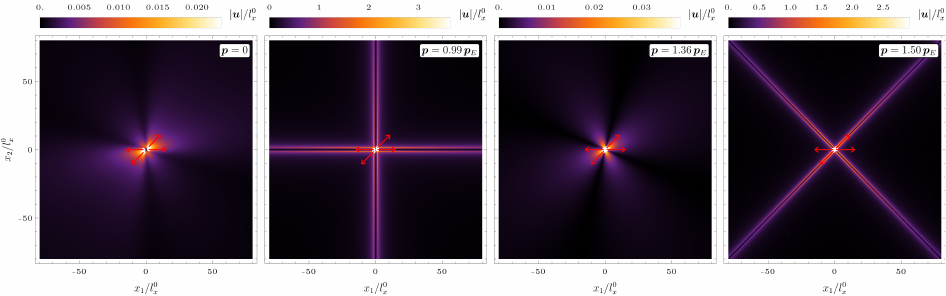}
    \caption{As for Fig.~\ref{figata}, except that the path labelled \lq 2' (satisfies a 45$^\circ$ symmetry) shown in Fig.~\ref{fig:stability_domains_cubic_deform} is considered and the perturbing agent is a quadrupole (shown red). Along a radial path entirely within the (SE) regime, a perturbation leads to the appearance of two shear bands (vertical and horizontal), their subsequent disappearance, and the final emergence of two shear bands (inclined at $45^\circ$ and $135^\circ$).}
    \label{figata_2}
\end{figure}

The model of axially and flexurally deformable elastic rods leads to the unexpected emergence of islands of instability in a region of stable (SE) behaviour. The presence of the islands implies that radial paths, such as those labelled \lq 2' and \lq 5' in Fig.~\ref{fig:stability_domains_cubic_deform}, can be defined, remaining inside (SE) but grazing its boundary and finally crossing it.
In this condition, a perturbation applied when the material is sufficiently close to the (SE) boundary reveals the occurrence of strain localization, which appears two times, the first when \lq grazing the island' and the second \lq in the proximity of the final frontier' (inclined at 45$^\circ$). This is illustrated in Figs.~\ref{figata} and \ref{figata_2}, referred to the paths labelled \lq 5' and \lq 2', respectively. The figures report the maps of (the modulus of) incremental displacements produced by the application of a perturbing force dipole (a quadrupole in Fig.~\ref{figata_2}) at increasing prestress, referred to the value $\bp_E$, corresponding to the first (SE) loss. In both figures, a dipole (obtained from the Green's function as explained in \cite{bordiga_dynamics_2021}) is applied in the equivalent material at a fixed position and inclined at $45^\circ$ (a horizontal dipole is also added to generate the quadrupole) in the undeformed configuration, so that, due to the effect of the prestrain, both the distance between the two forces and the inclination change (as visible in Fig.~\ref{figata}, while in Fig.~\ref{figata_2} this difference is not visible, because of the imposed symmetry of the prestress). 
%

Both figures show the appearance, disappearance, and reappearance of shear bands. In particular, Fig.~\ref{figata} shows the emergence of a first, \lq grazing the island', vertical shear band, corresponding to the SE/Parabolic boundary, while the second localization occurs in a mixed mode, becoming visible near the SE/Hyperbolic boundary, the \lq final frontier'. Fig.~\ref{figata_2} satisfies a $45^\circ$--symmetry, so that twin shear bands are always generated, horizontally and vertically the first time, inclined at $45^\circ$ the second time.

\section{Conclusions}
\label{sec:concluding}

This study introduces a conceptual model based on a unit cell with 10 degrees of freedom, which generates a rectangular lattice when periodically repeated to tessellate the entire plane. The unit cell consists of bars with only axial compliance, connected by elastic hinges. 
The proposed mechanical model enables a manual homogenization process to rigorously derive an equivalent elastic material.
The model reveals unexpectedly complex behaviour, in which strain localization may occur in the form of shear, compaction, or mixed-mode bands. Perhaps more interestingly, both the homogenized material and the underlying lattice demonstrate restabilization. Along a compressive radial path defined in the stability map, the material typically transitions from an initially stable behaviour to a loss of positive definiteness of the elasticity tensor, followed by ellipticity failure and shear band formation. As the compressive load increases along the path, the material traverses an unstable region, eventually restabilizing by entering a new stable region and culminating in a second loss of ellipticity and the emergence of mixed-mode strain localization. 
The mechanical behaviour observed in the lumped model has been confirmed through rigorous homogenization of a lattice composed of rods deformable in both bending and axial modes. This lattice exhibits additional remarkable features, such as closed instability domains within stable regions and the possibility of shear band formation and subsequent restabilization along a radial stress path entirely within the ellipticity domain. 

Despite their mechanical simplicity, these models exhibit a wide range of mechanical behaviours, including several novel features. This versatility can inspire a new generation of architected materials with tuneable instabilities.

\section*{Acknowledgements}
DB and AP acknowledge financial support from the European Research Council (ERC) under the European Union’s Horizon Europe research and innovation programme, Grant Agreement No. ERC-ADG-2021-101052956-BEYOND. The methodologies developed in this work fall within the aims of the GNFM (Gruppo Nazionale per la Fisica Matematica) of the INDAM (Istituto Nazionale di Alta Matematica).

\printbibliography

@article{ariza_homogenization_2024,
	title = {Homogenization and continuum limit of mechanical metamaterials},
	volume = {196},
	issn = {01676636},
	url = {https://linkinghub.elsevier.com/retrieve/pii/S0167663624001650},
	doi = {10.1016/j.mechmat.2024.105073},
	pages = {105073},
	journaltitle = {Mechanics of Materials},
	shortjournal = {Mechanics of Materials},
	author = {Ariza, M.P. and Conti, S. and Ortiz, M.},
	urldate = {2025-03-13},
	date = {2024-09},
	langid = {english},
}

@article{tarnai_destabilizing_1980,
	title = {Destabilizing effect of additional restraint on elastic bar structures},
	volume = {22},
	issn = {00207403},
	url = {https://linkinghub.elsevier.com/retrieve/pii/0020740380900545},
	doi = {10.1016/0020-7403(80)90054-5},
	pages = {379--390},
	number = {6},
	journaltitle = {International Journal of Mechanical Sciences},
	shortjournal = {International Journal of Mechanical Sciences},
	author = {Tarnai, T.},
	urldate = {2023-08-21},
	date = {1980-01},
	langid = {english},
}

@article{vanhove,
	title = {Sur l’extension de la condition de Legendre du calcul des variations aux int\'egrales multiples \`a plusieurs fonctions inconnues},
	volume = {50},
	pages = {18--23},
	journaltitle = {Proc. Sect. Sci. K. Akad. van Wetenschappen},
	shortjournal = {Proc. Sect. Sci. K. Akad. van Wetenschappen},
	author = {van Hove, L.},
	date = {1947},
	langid = {french},
}

@book{bigoni_nonlinear_2012,
	edition = {1},
	title = {Nonlinear Solid Mechanics: Bifurcation Theory and Material Instability},
	isbn = {9781107025417 9781139178938 9781107699502},
	url = {https://www.cambridge.org/core/product/identifier/9781139178938/type/book},
	shorttitle = {Nonlinear Solid Mechanics},
	publisher = {Cambridge University Press},
	author = {Bigoni, Davide},
	urldate = {2023-08-04},
	date = {2012-07-30},
	doi = {10.1017/CBO9781139178938},
}

@incollection{potier-ferry_foundations_1987,
	location = {Berlin/Heidelberg},
	title = {Foundations of elastic postbuckling theory},
	volume = {288},
	isbn = {9783540183129},
	url = {http://link.springer.com/10.1007/BFb0009197},
	pages = {1--82},
	booktitle = {Buckling and Post-Buckling},
	publisher = {Springer-Verlag},
	author = {Potier-Ferry, Michel},
	urldate = {2023-08-21},
	date = {1987},
	langid = {english},
	doi = {10.1007/BFb0009197},
}

@article{koutsogiannakis_double_2023,
	title = {Double restabilization and design of force–displacement response of the extensible elastica with movable constraints},
	volume = {100},
	issn = {09977538},
	url = {https://linkinghub.elsevier.com/retrieve/pii/S0997753822001887},
	doi = {10.1016/j.euromechsol.2022.104745},
	pages = {104745},
	journaltitle = {European Journal of Mechanics - A/Solids},
	shortjournal = {European Journal of Mechanics - A/Solids},
	author = {Koutsogiannakis, P. and Bigoni, D. and Dal Corso, F.},
	urldate = {2023-08-21},
	date = {2023-07},
	langid = {english},
}

@article{bigoni_instability_2014,
	title = {Instability of a penetrating blade},
	volume = {64},
	issn = {00225096},
	url = {https://linkinghub.elsevier.com/retrieve/pii/S0022509613002561},
	doi = {10.1016/j.jmps.2013.12.008},
	pages = {411--425},
	journaltitle = {Journal of the Mechanics and Physics of Solids},
	shortjournal = {Journal of the Mechanics and Physics of Solids},
	author = {Bigoni, D. and Bosi, F. and Dal Corso, F. and Misseroni, D.},
	urldate = {2023-08-21},
	date = {2014-03},
	langid = {english},
}

@incollection{pellegrino_deployable_2001,
	location = {Vienna},
	title = {Deployable Structures in Engineering},
	isbn = {9783211836859 9783709125847},
	url = {http://link.springer.com/10.1007/978-3-7091-2584-7_1},
	pages = {1--35},
	booktitle = {Deployable Structures},
	publisher = {Springer Vienna},
	author = {Pellegrino, Sergio},
	editor = {Pellegrino, S.},
	urldate = {2023-08-21},
	date = {2001},
	doi = {10.1007/978-3-7091-2584-7_1},
}

@article{bosi_asymptotic_2016,
	title = {Asymptotic self-restabilization of a continuous elastic structure},
	volume = {94},
	issn = {2470-0045, 2470-0053},
	url = {https://link.aps.org/doi/10.1103/PhysRevE.94.063005},
	doi = {10.1103/PhysRevE.94.063005},
	pages = {063005},
	number = {6},
	journaltitle = {Physical Review E},
	shortjournal = {Phys. Rev. E},
	author = {Bosi, F. and Misseroni, D. and Dal Corso, F. and Neukirch, S. and Bigoni, D.},
	urldate = {2023-08-21},
	date = {2016-12-27},
	langid = {english},
}

@book{feodosyev_selected_1977,
	title = {Selected Problems and Questions in Strength of Materials},
	isbn = {978-0-8464-1436-0},
	url = {https://books.google.it/books?id=kB3QngEACAAJ},
	publisher = {Mir Publishers},
	author = {Feodosyev, V.I.},
	date = {1977},
    address = {{Moscow}},
	lccn = {79345546},
}

@article{bordiga_dynamics_2021,
	title = {Dynamics of prestressed elastic lattices: Homogenization, instabilities, and strain localization},
	volume = {146},
	issn = {00225096},
	url = {https://linkinghub.elsevier.com/retrieve/pii/S0022509620304208},
	doi = {10.1016/j.jmps.2020.104198},
	shorttitle = {Dynamics of prestressed elastic lattices},
	pages = {104198},
	journaltitle = {Journal of the Mechanics and Physics of Solids},
	shortjournal = {Journal of the Mechanics and Physics of Solids},
	author = {Bordiga, G. and Cabras, L. and Piccolroaz, A. and Bigoni, D.},
	urldate = {2024-10-21},
	date = {2021-01},
	langid = {english},
}

@article{bordiga_tensile_2022,
	title = {Tensile material instabilities in elastic beam lattices lead to a bounded stability domain},
	volume = {380},
	issn = {1364-503X, 1471-2962},
	url = {https://royalsocietypublishing.org/doi/10.1098/rsta.2021.0388},
	doi = {10.1098/rsta.2021.0388},
	pages = {20210388},
	number = {2231},
	journaltitle = {Philosophical Transactions of the Royal Society A: Mathematical, Physical and Engineering Sciences},
	shortjournal = {Phil. Trans. R. Soc. A.},
	author = {Bordiga, Giovanni and Bigoni, Davide and Piccolroaz, Andrea},
	urldate = {2024-10-21},
	date = {2022-09-05},
	langid = {english},
}

@article{lopez-pamies_overall_2006,
	title = {On the overall behavior, microstructure evolution, and macroscopic stability in reinforced rubbers at large deformations: I—Theory},
	volume = {54},
	rights = {https://www.elsevier.com/tdm/userlicense/1.0/},
	issn = {00225096},
	url = {https://linkinghub.elsevier.com/retrieve/pii/S0022509605002024},
	doi = {10.1016/j.jmps.2005.10.006},
	shorttitle = {On the overall behavior, microstructure evolution, and macroscopic stability in reinforced rubbers at large deformations},
	pages = {807--830},
	number = {4},
	journaltitle = {Journal of the Mechanics and Physics of Solids},
	shortjournal = {Journal of the Mechanics and Physics of Solids},
	author = {Lopez-Pamies, O. and Ponte Castañeda, P.},
	urldate = {2024-11-18},
	date = {2006-04},
	langid = {english},
}

@article{triantafyllidis_comparison_1993,
	title = {Comparison of microscopic and macroscopic instabilities in a class of two-dimensional periodic composites},
	volume = {41},
	rights = {https://www.elsevier.com/tdm/userlicense/1.0/},
	issn = {00225096},
	url = {https://linkinghub.elsevier.com/retrieve/pii/002250969390039I},
	doi = {10.1016/0022-5096(93)90039-I},
	pages = {1533--1565},
	number = {9},
	journaltitle = {Journal of the Mechanics and Physics of Solids},
	shortjournal = {Journal of the Mechanics and Physics of Solids},
	author = {Triantafyllidis, Nicolas and Schnaidt, William C.},
	urldate = {2024-11-18},
	date = {1993-09},
	langid = {english},
}

@article{geymonat_homogenization_1993,
	title = {Homogenization of nonlinearly elastic materials, microscopic bifurcation and macroscopic loss of rank-one convexity},
	volume = {122},
	rights = {http://www.springer.com/tdm},
	issn = {0003-9527, 1432-0673},
	url = {http://link.springer.com/10.1007/BF00380256},
	doi = {10.1007/BF00380256},
	pages = {231--290},
	number = {3},
	journaltitle = {Archive for Rational Mechanics and Analysis},
	shortjournal = {Arch. Rational Mech. Anal.},
	author = {Geymonat, Giuseppe and M{\"u}ller, Stefan and Triantafyllidis, Nicolas},
	urldate = {2024-11-18},
	date = {1993},
	langid = {english},
}

@article{santisi_davila_localization_2016,
	title = {Localization of deformation and loss of macroscopic ellipticity in microstructured solids},
	volume = {97},
	issn = {00225096},
	url = {https://linkinghub.elsevier.com/retrieve/pii/S0022509616304677},
	doi = {10.1016/j.jmps.2016.07.009},
	pages = {275--298},
	journaltitle = {Journal of the Mechanics and Physics of Solids},
	shortjournal = {Journal of the Mechanics and Physics of Solids},
	author = {Santisi d'Avila, M.P. and Triantafyllidis, N. and Wen, G.},
	urldate = {2024-12-09},
	date = {2016-12},
	langid = {english},
}

@article{michel_microscopic_2010,
	title = {Microscopic and macroscopic instabilities in finitely strained fiber-reinforced elastomers},
	volume = {58},
	rights = {https://www.elsevier.com/tdm/userlicense/1.0/},
	issn = {00225096},
	url = {https://linkinghub.elsevier.com/retrieve/pii/S0022509610001687},
	doi = {10.1016/j.jmps.2010.08.006},
	pages = {1776--1803},
	number = {11},
	journaltitle = {Journal of the Mechanics and Physics of Solids},
	shortjournal = {Journal of the Mechanics and Physics of Solids},
	author = {Michel, J.C. and Lopez Pamies, O. and Ponte Castaneda, P. and Triantafyllidis, N.},
	urldate = {2024-12-09},
	date = {2010-11},
	langid = {english},
}

@article{michel_microscopic_2007,
	title = {Microscopic and macroscopic instabilities in finitely strained porous elastomers},
	volume = {55},
	rights = {https://www.elsevier.com/tdm/userlicense/1.0/},
	issn = {00225096},
	url = {https://linkinghub.elsevier.com/retrieve/pii/S0022509606001815},
	doi = {10.1016/j.jmps.2006.11.006},
	pages = {900--938},
	number = {5},
	journaltitle = {Journal of the Mechanics and Physics of Solids},
	shortjournal = {Journal of the Mechanics and Physics of Solids},
	author = {Michel, J. and Lopez Pamies, O. and Ponte Castaneda, P. and Triantafyllidis, N.},
	urldate = {2024-12-09},
	date = {2007-05},
	langid = {english},
}


%

\end{document}